\documentclass[a4paper,fleqn,usenatbib]{style/mnras}

\usepackage{newtxtext,newtxmath}

\usepackage[T1]{fontenc}
\usepackage{ae,aecompl}

\usepackage{times}
\usepackage[none]{hyphenat}

\usepackage{graphicx}	
\usepackage{amsmath}	
\usepackage{amssymb}	
\usepackage{setspace}
\usepackage{graphics}
\usepackage{graphicx}
\usepackage{longtable}
\usepackage{amsbsy,fixmath}
\usepackage{epstopdf}
\usepackage{color}
\usepackage[T1]{fontenc}
\usepackage{aecompl}
\usepackage[normalem]{ulem}
\usepackage{physics}
\usepackage{mathrsfs}

\bibpunct[,]{(}{)}{;}{a}{}{,}

\usepackage{colortbl}
\usepackage{xcolor}
\usepackage{bm}
\usepackage{float}
\usepackage{multicol}

\usepackage{tabularx}

\usepackage{csquotes}
\usepackage{multirow}
\usepackage{rotating}

\newcommand{\arepo}{\textsc{arepo}}

\newcommand{\GRR}{\citet{Guillochon2013}}
\newcommand{\pGRR}{\citep{Guillochon2013}}

\title[Moving-mesh simulations of tidal disruption events]{Hydrodynamical moving-mesh
  simulations of the tidal disruption of stars by supermassive black holes}

\author[F. G. Goicovic et al.]{
Felipe~G.~Goicovic,$^{1,2}$\thanks{E-mail: felipe.goicovic@gmail.com (FGG)}
Volker~Springel,$^{3}$
Sebastian~T.~Ohlmann$^{4,1}$ and
R\"udiger~Pakmor$^{3}$
\\
$^{1}$Heidelberg Institute for Theoretical Studies (HITS), Schloss-Wolfsbrunnenweg 35, D-69118 Heidelberg, Germany\\
$^{2}$Zentrum f\"ur Astronomie der Universit\"at Heidelberg, Institut f\"ur Theoretische Astrophysik, Albert-\"Uberle-Str.~2, 69120 Heildelberg, Germany\\
$^{3}$Max-Planck-Institut f\"ur Astrophysik, Karl-Schwarzschild-Str. 1, D-85748 Garching, Germany\\
$^{4}$Max Planck Computing and Data Facility, Gie\ss{}enbachstr. 2, D-85748 Garching, Germany
}

\date{\today}

\pubyear{2018}

\begin{document}
\label{firstpage}
\pagerange{\pageref{firstpage}--\pageref{lastpage}}
\maketitle

\begin{abstract}
  When a star approaches a black hole closely, it may be pulled apart
  by gravitational forces in a tidal disruption event (TDE).  The
  flares produced by TDEs are unique tracers of otherwise quiescent
  supermassive black holes (SMBHs) located at the centre of most
  galaxies.  In particular, the appearance of such flares and the
  subsequent decay of the light curve are both sensitive to whether
  the star is partially or totally destroyed by the tidal
  field. However, the physics of the disruption and the fall-back of
  the debris are still poorly understood.  We are here modelling the
  hydrodynamical evolution of realistic stars as they approach a SMBH
  on parabolic orbits, using for the first time the moving-mesh code
  \arepo, which is particularly well adapted to the problem through
  its combination of quasi-Lagrangian behaviour, low advection errors,
  and high accuracy typical of mesh-based techniques.  We examine a
  suite of simulations with different impact parameters, allowing us
  to determine the critical distance at which the star is totally
  disrupted, the energy distribution and the fallback rate of the
  debris, as well as the hydrodynamical evolution of the stellar
  remnant in the case of a partial disruption.  Interestingly, we find
  that the internal evolution of the remnant's core is strongly
  influenced by persistent vortices excited in the tidal
  interaction. These should be sites of strong magnetic field
  amplification, and the associated mixing may profoundly alter the
  subsequent evolution of the tidally pruned star.
\end{abstract}




\section{Introduction}

Supermassive black holes (SMBHs) have been observed in the centre of
most massive galaxies \citep[e.g.][]{Ferrarese2005}, and are believed
to play a fundamental role in galaxy evolution
\citep[e.g.][]{DiMatteo2005}.  Most notably, by accreting gas from
their surroundings, these objects are capable of emitting enormous
amounts of energy.  Unfortunately, actively accreting black holes (or
AGNs) represent only a small fraction of the entire supermassive black
hole population, making it challenging to observe them through most of
their life time.

Alternatively, because SMBHs are usually embedded within dense stellar
clusters, the disruption of stars can provide material to power short
periods of activity \citep{Frank1976}.  When a star passes too close
to a black hole, the tidal forces overcome its self-gravity, which
tears the star apart.  A fraction of the stripped stellar material
remains bound to the SMBH, eventually forming an accretion disc and
powering activity that can last from months to years \citep{Rees1988},
with peak values even comparable to the Eddington luminosity.  These
events are often referred to as tidal disruption events (TDEs), and
they constitute a powerful indirect probe for studying black holes in
the centre of galaxies, both in the local and distant Universe.

The theoretical basis to understand the emission from TDEs was laid
down during the eighties in seminal works by \citet{Lacy1982},
\citet{Rees1988}, \citet{Phinney1989}, and from a numerical
perspective, \citet{Evans1989}.  Their models showed that TDEs were
detectable from UV to soft X-ray wavelengths with a light curve
decreasing characteristically as $t^{-5/3}$.  TDE candidates
were later observed with properties in broad agreement with these
theoretical predictions \citep[e.g.][]{Bade1996,Komossa1999,
  Gezari2006,Komossa2008, vanVelzen2011}, strengthening this
theoretical background.  However, detailed observations of
different TDE candidates have also found signatures that cannot be
explained by the standard model \citep[see the illustrative cases
presented in][]{Bloom2011,Gezari2012}, suggesting that the picture
is more complex than originally modelled.

The tidal disruption of a star and the subsequent fallback of gas is
primarily governed by gravity, hence the basic principles of TDEs are
well understood. However, in reality there are several additional
physical processes at play, making the accurate analytical (or
semi-analytical) modelling of these events quite challenging.
Numerical hydrodynamical simulations are therefore the tool of choice
for more detailed calculations.  There has been significant progress
on this front in recent years, helped also by the development of an
increasing variety of suitable hydrodynamical codes.  TDE simulations
typically used either the Lagrangian smoothed particle hydrodynamics
(SPH) technique \citep[e.g.][]{Lodato2009, Rosswog2009, Tejeda2013,
  Coughlin2015, Bonnerot2016, Coughlin2016, Sadowski2016,
  Mainetti2017}, or Eulerian grid-based methods
\citep[e.g.][]{Evans1989,Khokhlov1993,Khokhlov1993b,
  Frolov1994,Diener1997,Guillochon2009,Guillochon2013,Cheng2014}.
Although both of these approaches have their particular advantages,
they also have some important limitations.  Grid-based codes, for
example, are not manifestly Galilean invariant and do not conserve
angular momentum, which is crucial to accurately follow orbits.  On
the other hand, SPH methods tend to significantly broaden shocks, are
associated with numerical surface tension effects which suppress
mixing, and are comparatively noisy. These numerical deficits could
introduce significant inaccuracies in simulations of TDEs, either
during the disruption itself, the subsequent fallback of stripped
material, or the hydrodynamical evolution of a partially disrupted
star.

In recent years, new simulation methods have been developed with the
goal of combining the advantages of both SPH and grid-based techniques
while avoiding some of their disadvantages. In particular, new
quasi-Lagrangian algorithms where the volume is discretised using a
set of mesh-generating tracers that move with the fluid, as pioneered
in the hydrodynamical code \arepo~\citep{Springel2010}, have proven to
be a robust and versatile tool to model a large variety of
astrophysical systems \citep[e.g.][]{Greif2011, Vogelsberger2012,
  Nelson2013, Marinacci2014, Zhu2015, Ohlmann2016, Weinberger2017,
  Springel2018}. In particular, because the mesh moves with the gas,
highly supersonic flows do not suffer accuracy degradation due to
advection errors in this approach, while the good shock-capturing
accuracy and the ability to follow turbulence of ordinary mesh-based
techniques are retained. These features are ideal for the TDE problem.

In this paper, we present a suite of simulations of the disruption of
zero-age-main-sequence stars by a SMBH, from the first approach until
several hours after periapsis, using \arepo.  Our models for the first
time apply the moving-mesh technique to TDEs, and also represent the
first examples of this type of simulations based on realistic stellar
structure profiles.  Previously, the structure of main sequence stars
in tidal disruption simulations has been modelled almost exclusively
using single polytropes. While this can be a decent approximation
in many cases, it does not allow for more evolved stars that tend to
be more centrally concentrated.

The paper is organised as follows. We describe the numerical methods
and setup of our models in Section~\ref{sec:simulations}.  In
Section~\ref{sec:results}, we present a determination of the critical
distance at which the star is completely destroyed, and the
corresponding energy distribution and fallback rate of the debris.  In
Section~\ref{sec:core}, we study the evolution of the surviving
stellar core after a partial disruption.  We finally summarise and
discuss our results in Section~\ref{sec:conclusions}.

\section{Numerical methods}
\label{sec:simulations}

We simulate the close encounter between a star and a single black hole
using the finite volume hydrodynamics code
\arepo~\citep{Springel2010,Pakmor2016}.  This code solves the Euler
equations using a finite-volume approach on an unstructured Voronoi
mesh that is generated from a set of points that move with the
flow. The mesh admits the application of second-order accurate Godunov
methods for evolving the fluid state in time, similar to the ones
known from standard Eulerian finite volume hydrodynamical
codes. However, the fluxes across cell boundaries are solved in the
moving frame of mesh faces, which have minimal residual motion with
respect to the fluid itself. This greatly diminishes advection errors
inherent in ordinary Eulerian treatments, and prevents that the
\arepo~results degrade in their accuracy with increasing bulk velocity
of the star. At the same time, the use of a second-order accurate
reconstruction yields high spatial and temporal accuracy, yielding a
considerably faster convergence rate than achieved in SPH, where
numerical noise and errors in discrete kernel sums cause much slower
convergence speeds. Also, our method does not need to impose an
artificial viscosity and naturally resolves mixing that may happen in
a multidimensional flow.

Another advantage of the quasi-Lagrangian nature of \arepo~lies in its
automatic adaptivity to the flow, allowing the spatial resolution to
smoothly and continuously adjust to the mass density without imposing
preferred grid directions. The full spatial and temporal adaptivity of the
scheme is further strengthened by the ability to refine or derefine
cells as needed, similar to how this is done in adaptive mesh
refinement codes. We use these features to maintain constant mass
resolution within the star and its stripped material, while
guaranteeing a minimum spatial resolution in low density regions
throughout the computational domain. Taken together, we think this
method is hence particularly well suited for the TDE problem.

\begin{figure}
\centering
\includegraphics[width=\columnwidth]{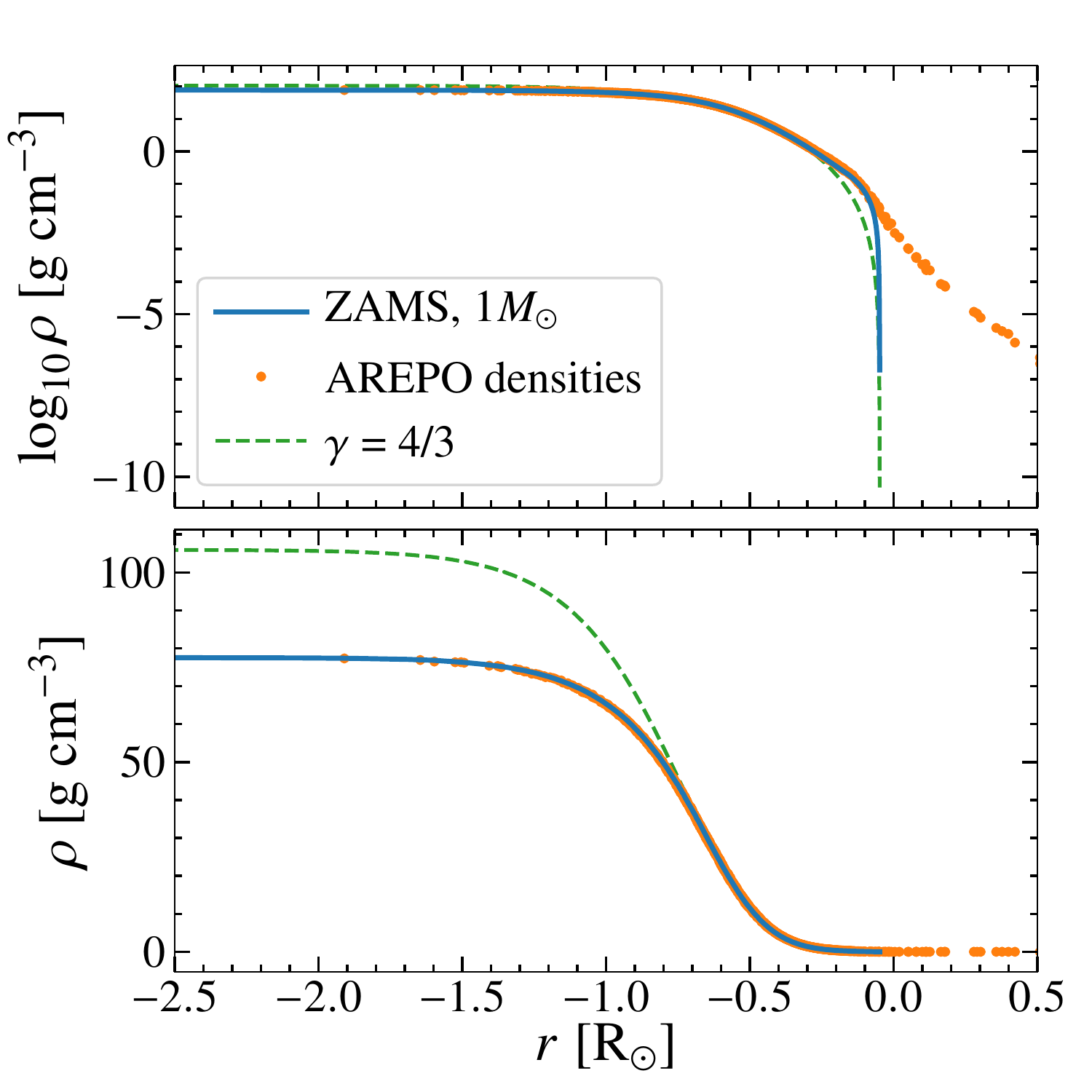}
\caption{Density profiles of a $1\,{\rm M}_\odot$ ZAMS star.  The
  {light blue} solid line shows the one dimensional profile from MESA,
  while the {orange} points display the individual densities computed by
  \arepo~after the relaxation run.
  {For clarity the latter is shown only for a subset of 2000 randomly
  selected gas cells.}
  For comparison, the {green} dashed line shows a polytropic
  profile with index $\gamma=4/3$, and the same total mass and radius
  of the 1D profile.  This polytropic profile is often used to
  represent high-mass stars.}
\label{fig:profiles}
\end{figure}

The stellar model we use in our tidal disruption simulations is created
with the help of the stellar evolution code MESA \citep[Modules for
Experiments in Stellar Astrophysics;][]{Paxton2011,Paxton2013}, version
7623. We create a zero-age main sequence (ZAMS) star of mass $1\,{\rm M}_\odot$
and metallicity 0.02 as an input model for the hydrodynamics simulations.

\begin{figure*}
\centering
\includegraphics[width=\textwidth]{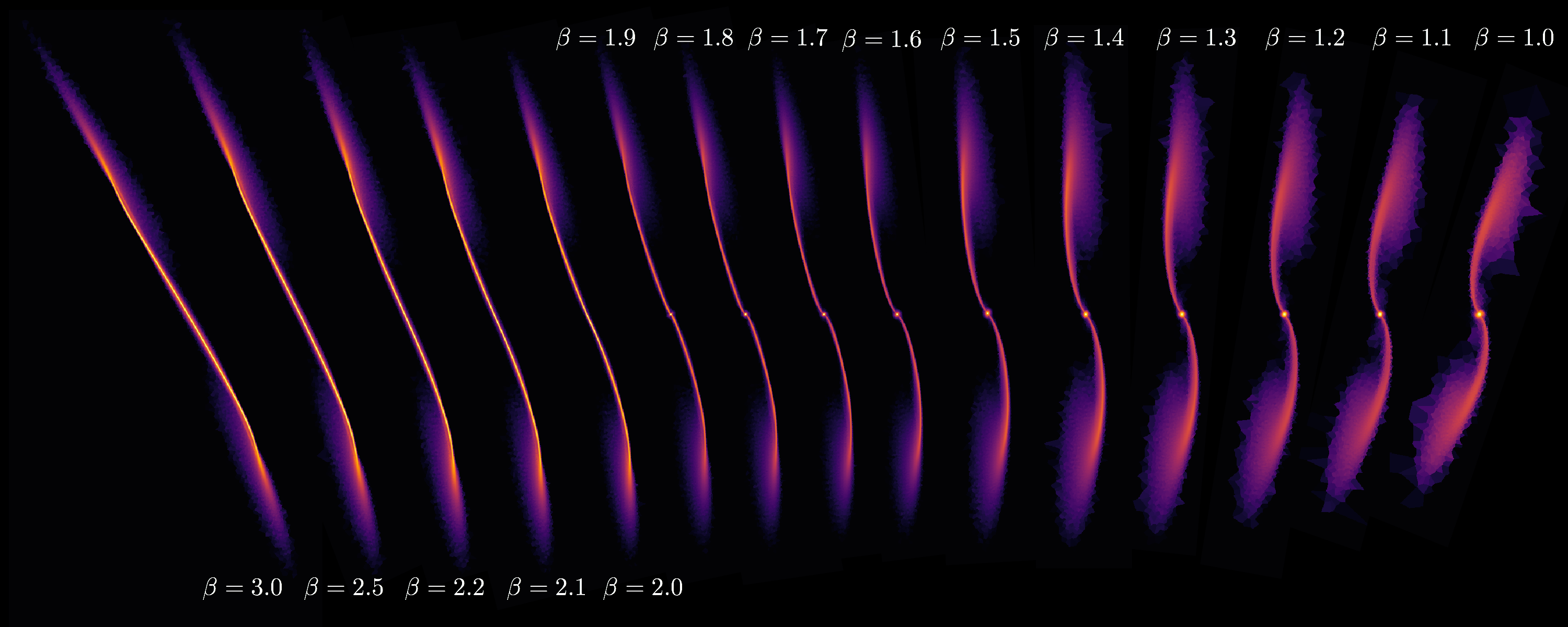}
\caption{Total column density (in logarithmic scale) of each
  simulation's final output for the disruption of a $1\,{\rm M}_\odot$
  ZAMS star by a $10^6\,{\rm M}_\odot$ black hole.  Each output is
  obtained $1.5\times 10^{5}$ seconds after periapsis, and is labelled
  with the corresponding value of the penetration parameter $\beta$, with
  the deepest encounter starting from the left.  Here we observe the
  transition from the regime where the star is completely destroyed
  for deep encounters ($\beta\gtrsim 2$), to a small surviving core
  for smaller penetration parameters.}
\label{fig:rho}
\end{figure*}

To produce the actual three-dimensional initial distribution of gas
cells we follow a procedure fully described in \citet{Ohlmann2017}
that we briefly summarise here.  First, the one-dimensional stellar
profile from MESA is mapped to a 3D grid using concentric shells with
a HEALPix angular distribution on each shell.  We place this spherical
distribution into a small periodic box with $4\,R_\odot$ on a side and
a very low background density of $10^{-16}$ g cm$^{-3}$.  This
configuration is then relaxed using a damping procedure over a time
$10\,t_{\rm dyn}$, where $t_{\rm dyn}$ is the sound crossing time of the
star.  This is done to eliminate spurious velocities resulting from
the discretization on our mesh, and results in a stable profile
according to the stability criteria defined in \citet{Ohlmann2017}.

We show the resulting stellar density profile in
Fig.~\ref{fig:profiles}. The {light blue} solid line is the spherically
symmetric profile from MESA, while the {orange} points
correspond to the individual densities computed by \arepo~at the
end of the relaxation run.
We treat the gas as ideal, with an adiabatic equation of state and
an adiabatic index of $5/3$.
Note that the stellar profile after relaxation follows very closely the
initial MESA profile at most radii, with the exception of the stellar
surface where the initial density profile drops to sharply.
This discrepancy arises because a significantly
larger number of cells would be necessary to more accurately resolve
the steep gradient at the stellar surface.  As this region contains a
negligible fraction of the total stellar mass, this is not
necessary in our application and this disagreement does not affect our
results.

We additionally show a polytropic profile with index $\gamma=4/3$ in
Fig.~\ref{fig:profiles} ({green} dashed line), which is often used to
represent high-mass stars in this type of simulations.  This profile
is not dramatically different from the one obtained from MESA, only
slightly more centrally concentrated, and arguably not a bad
approximation for our star.  However, the strength of the procedure
presented in this paper is that it can be extended to \textit{any}
stage of stellar evolution, which opens up the possibility of
modelling tidal disruptions for a whole suite of different stars, with
varying mass and age, but always using a realistic internal structure. This
includes giant stars, for example, where the core could additionally
be replaced by a point mass to represent the extreme density contrast
between the core and the envelope \citep{Ohlmann2017} in an efficient
fashion.  As demonstrated by \GRR, the stellar structure is crucial
for determining the characteristics of the disruption (e.g. the
critical distance for total disruption and the energy distribution of
the debris). Consequently, it is of paramount importance to use a
physically motivated structure for the stars in TDE simulations to
ultimately produce realistic light curve predictions.

Once we have the relaxed stellar profile, we can proceed to model the
star's tidal disruption by the black hole.  In our simulations, the
SMBH is simply modelled as an external Newtonian point potential,
located at the centre of the domain, with a total mass of
$M_{\rm BH}=10^6\,{\rm M}_\odot$.  One of the most relevant scales to
take into account when modelling TDEs is the distance at which the
tidal forces of the black hole are larger than the star's gravity at
its surface, which is referred to as the tidal radius
\begin{equation}
r_t\equiv R_*\qty(\frac{M_{\rm BH}}{M_*})^{1/3} \simeq
7\times 10^{12}\,\mbox{cm}\,\qty(\frac{M_{\rm BH}}{10^6M_\odot})
\qty(\frac{R_*}{R_\odot})\qty(\frac{M_*}{M_\odot})^{-1/3},
\end{equation}
where $R_*$ and $M_*$ denote the radius and mass of the star,
respectively, and $M_{\rm BH}$ is the mass of the black hole.  Inside
this sphere the star's gravity can no longer prevent material from
being stripped, and the star is disrupted at least partially.  With
our choice of black hole mass, the {black hole to stellar mass ratio}
corresponds to $q=10^{-6}$, which results in a tidal radius of
$r_t = 100\,R_*$.

The Newtonian approximation for the black hole is still reasonable for
our models, since we are simulating encounters were the closest
approach is $\sim 15\,r_g$, where $r_g=GM_{\rm BH}/c^2$ is the
gravitational radius of the black hole.  The relativistic corrections
during periapsis in this regime are expected to be small
\citep{Cheng2014, Stone2019}.  A more accurate treatment is needed
during the fallback and accretion of the debris, since relativistic
effects are important for the formation of the accretion disc and its
evolution \citep[see e.g.][]{Bonnerot2016}.  This process, however, is
beyond the scope of the present work as we focus on the disruption
process itself.

To model the disruption, we place the relaxed star in a periodic box
with a side length of $2.1\times 10^{15}\,{\rm cm}$ (equivalent to
$300\, r_t$), with a low background density of
$10^{-16}\,{\rm g\, cm}^{-3}$.  All the simulations presented in this
paper were stopped after about 44 hours, with the leading arm of the
disrupted star still being far from the boundary, ensuring that there
is no stellar mass flowing over an edge of the domain and reappearing
from the opposite side.  Consequently, we do not have unphysical
effects due to the finite size of the box. We note that in contrast to
the hydrodynamics, the gravity of the BH and the self-gravity of the
stellar material are treated without periodic boundary conditions.

The initial velocity of the star is set up such that the star's centre of
mass describes a parabolic orbit, with the location of the periapsis
radius given by
\begin{equation}
r_{\rm p}=\frac{r_{\rm t}}{\beta},
\end{equation}
where $\beta$ is the so-called ``penetration'' parameter.  We vary
this parameter to obtain a suite of simulations with different stellar
orbits. In this paper, we present the results of 15 simulations, with
$\beta$ between 1 and 3. In each case, the star was placed at an
initial distance from the black hole equivalent to $5\,r_t$, which
ensures that the star's stability is preserved at the beginning of the
approach.  With this choice of parameters, the periapsis of the star
occurs between about 2 and 3 hours from the start of the simulation,
depending on the impact parameter.

Because we want the star to be represented by at least a total of
roughly $2\times 10^5$ cells, we set the refinement cell criterion to
a target cell mass of
$M_{\rm target}=9.8\times 10^{27}\,{\rm g} =4.9\times 10^{-6}\,{\rm
  M}_\odot$.  In addition to the mass refinement criteria, we use a
volume limit criterion in which neighbouring cells are refined such
that the ratio of their volumes is never larger than 5.  This allows
us to have more resolution elements in regions where there is little
mass, namely, in the outer layers of the star, and later on, in the
streams of stripped gas.  At the start of the simulations we have a
total of about $2.3\times 10^5$ cells, with $3\times 10^4$ forming
part of background grid, and a mean cell mass of
$8.9\times 10^{27}\,{\rm g}$.  By the end of the runs, the total
number of cells has increased to about $2.4\times 10^5$ with a mean
cell mass of $8.5\times 10^{27}\,{\rm g}$, thanks to the mesh
refinement.
To confirm the convergence of our results, we have also run a
subset of models with 10 times the resolution of our standard setup,
finding no appreciable difference. We thus only show the simulations
with the standard resolution explained above.

\section{{Influence of the penetration parameter}}
\label{sec:results}

\subsection{{The limit between total and partial disruption}}

As explained in the previous section, we simulate
stellar orbits with pericentre distances always smaller or
equal to the tidal radius. Contrary to intuition, however,
this does not mean that the star will be completely destroyed
in each of the simulated encounters, since the definition
of $r_t$ only considers the stellar gravity on its surface.
Whether the star is completely
or partially destroyed for a given impact parameter depends
almost exclusively on the internal stellar structure, with the
more centrally concentrated stars being able to survive deeper
encounters \citep[e.g.][]{Guillochon2013}.

In Figure~\ref{fig:rho}, we show density maps of our simulations. Each
output is obtained $1.5\times 10^{5}$ seconds after periapsis, which
corresponds to approximately 50 dynamical times.  In this figure, we
can observe the transition between total disruption for the deepest
encounters, to the point where the core starts re-collapsing for lower
values of $\beta$. Motivated by this result, we seek to identify the
critical distance of the star to the SMBH at which the encounter
results in a total disruption.

Since we do not simulate the encounter long enough for the core to
completely re-collapse, we compute the total stellar mass loss
$\Delta M$ at the end of our simulations by
using the method introduced by \GRR, in which the gas
gravitationally bound to the star is iteratively determined.  The
specific binding energy of each cell is computed as
\begin{equation}
E_{*,i}=\frac{1}{2}(\vb{v}_i-\vb{v}_*)^2 + \phi_i,
\end{equation}
where $\vb{v}_i$ is the cell's velocity, and $\phi_i$ is the
gravitational potential exerted by the rest of the gas onto each cell,
which is computed directly by \arepo~through its gravity solver. The
star's velocity $\vb{v}_*$ is initially chosen to be equal to the
velocity of the highest density peak.  After the first estimation of
$E_{*,i}$, we then consider only cells with $E_{*,i}<0$ to find a more
robust value for the star's location and velocity
\begin{equation}
\vb{r}_*=\frac{\sum_{E_{*,i}<0}m_i\vb{r}_i}{\sum_{E_{*,i}<0}m_i},
\end{equation}
\begin{equation}
\vb{v}_*=\frac{\sum_{E_{*,i}<0}m_i\vb{v}_i}{\sum_{E_{*,i}<0}m_i}.
\end{equation}
We iterate this procedure until the star's velocity has converged to a
constant value. Subsequently, the stellar mass loss is simply taken as
all the gas cells that are unbound from the star, i.e., which have
$E_{*,i}>0$.

\begin{figure}
\centering
\includegraphics[width=\columnwidth]{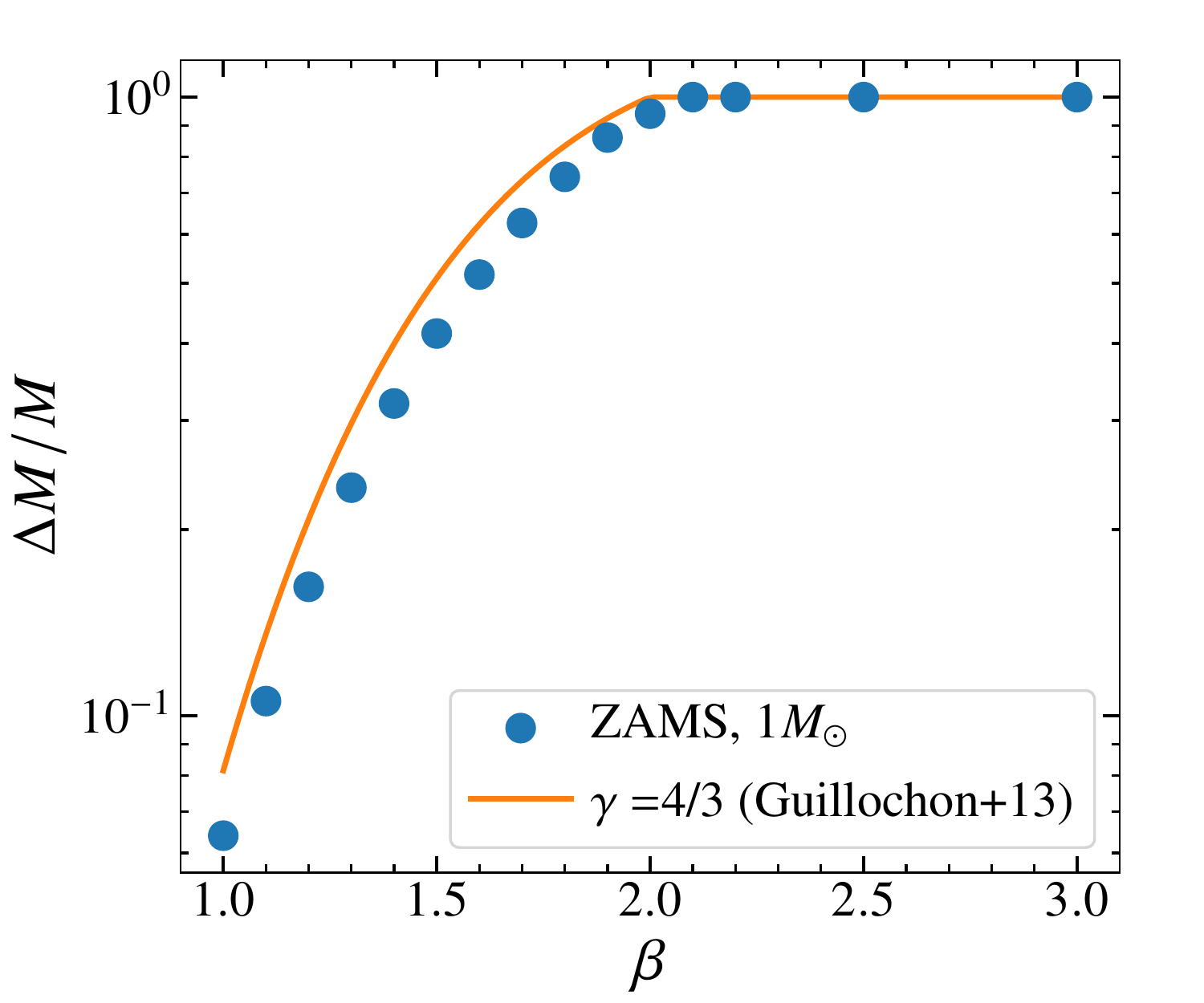}
\caption{Total mass that is tidally stripped from the star, in units
  of the initial mass, as a function of the penetration parameter.
  This quantity is measured at the end of the simulations.
  When this value reaches unity, full disruption has occurred.
  The solid line shows the fitting formula obtained by \GRR~for
  a $\gamma=4/3$ polytrope.  }
\label{fig:deltam}
\end{figure}

\begin{figure}
\centering
\includegraphics[width=\columnwidth]{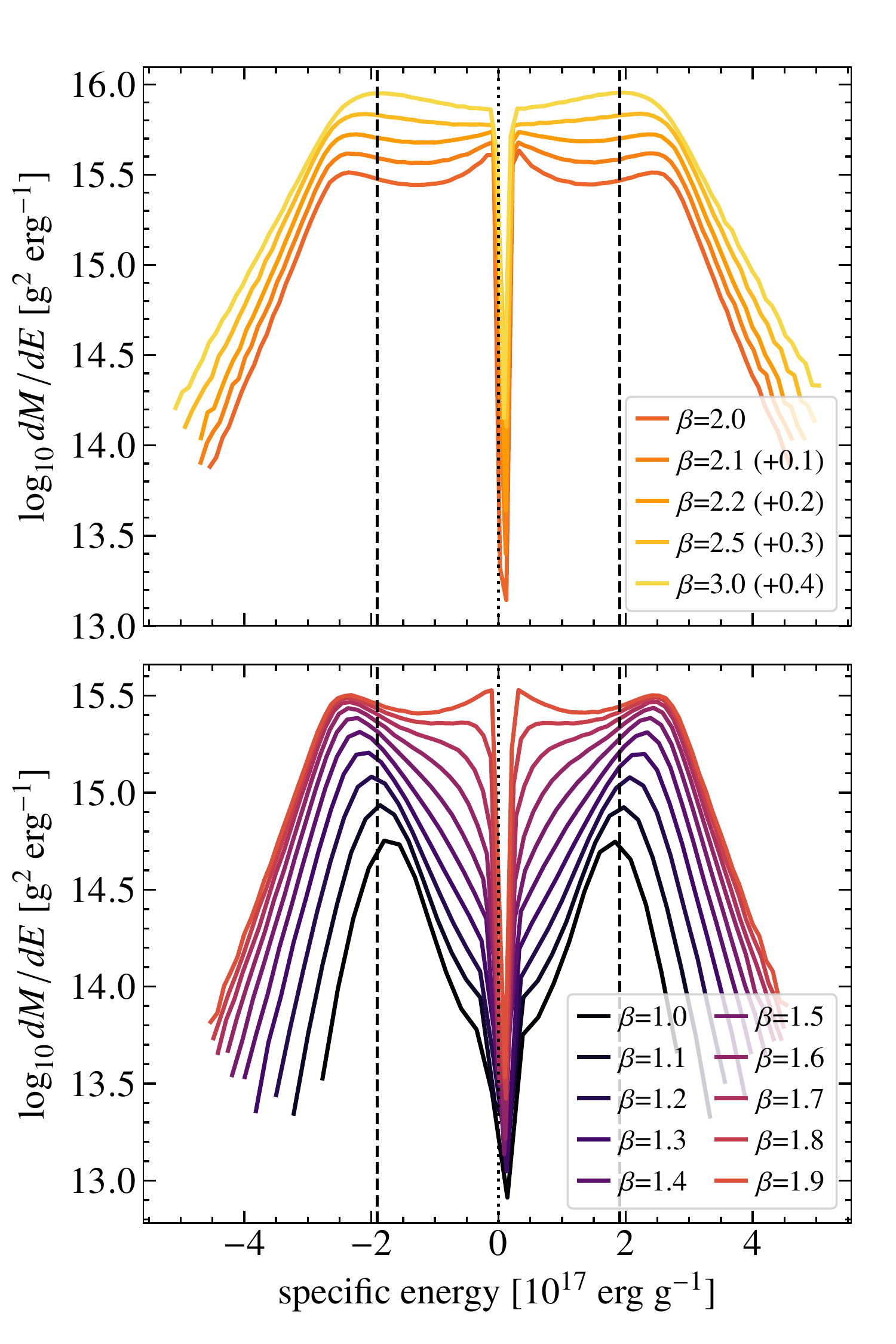}
\caption{Distribution of specific binding energy of the stripped
  material for different penetration parameters, increasing from the
  darkest to the lightest line colour. The vertical dashed lines
  indicate the expected energy spread from the `frozen in'
  approximation. This stripped material corresponds to the
  value of the total mass unbound from the star, as presented in
  Fig.~\ref{fig:deltam}, measured at the end of the simulation.
  {For clarity we separate the cases in which there is a surviving core
  ($\beta<2$, lower panel) and of total disruption ($\beta\geq 2$, upper panel).
  Furthermore, in the upper panel all the cases
  are arbitrarily shifted upwards with respect to $\beta=2$,
  and the relative shift is indicated in the legend.} }
\label{fig:energy}
\end{figure}

In Fig.~\ref{fig:deltam}, we show the total mass loss as a function of
the penetration parameter. Because $\Delta M$ is expressed in terms of
the star's initial mass, a total disruption occurs when this value reaches
unity.  We check the convergence of $\Delta M$ through the quantity
\begin{equation}
\mathcal{F}(t)\equiv \abs{\frac{\dot M_{\rm bound}}{M_{\rm bound}}}(t-t_p),
\label{eq:f}
\end{equation}
where $M_{\rm bound}$ is the total mass still bound to the star and
$t_{p}$ is the time of periapsis \pGRR.  We find the two expected
regimes: small and decreasing values of $\mathcal{F}$ when the core
survives, whereas $\mathcal{F}\sim 1$ at all times for total
disruptions.
{Notice that $M_{\rm bound}$ never formally reaches zero since
the tidal forces vanish towards the very center of the stellar remnant,
and thus this procedure always yields some gravitationally bound material,
even if the star is completely destroyed.
Nevertheless, following a total disruption, the bound gas quickly changes
with time as the material is perpetually stretched to never re-collapse,
resulting in $\mathcal{F}\sim 1$ at all times.}

In Fig.~\ref{fig:deltam}, we also show the fitting formula obtained by
\GRR~for the disruption of a $1\,{\rm M}_\odot$ star, represented with
a $\gamma=4/3$ polytrope.  As discussed in the previous section, this
profile has a similar shape as our ZAMS star, but is more centrally
concentrated.  As a consequence, the mass loss as a function of
$\beta$ in our simulations is quite similar to the one obtained with a
polytrope, including the value of the critical distance for total
disruption.  The star in our simulations looses slightly less mass
with respect to the polytrope, which is most likely a consequence of
the different concentrations, because our star has less mass than the
corresponding polytrope at a given radius in the inner regions.

Another effect that might cause different mass loss as a function of
impact parameter is the numerical technique. \citet{Mainetti2017}
numerically modelled the disruption of stars (represented by
polytropes) using different simulation techniques: mesh-free finite
mass, traditional SPH, and ``modern'' SPH.  They find that the
critical distance depends weakly on the method.  Considering that the
moving-mesh approach of \arepo~represents a different technique with
respect to the ones tested in \citet{Mainetti2017}, one expects
some differences to our results as well, although they are probably
subdominant with respect to the differences in the actual stellar
structure between our ZAMS star and a polytrope.

\begin{figure}
\begin{tabular}{c}
\includegraphics[width=\columnwidth]{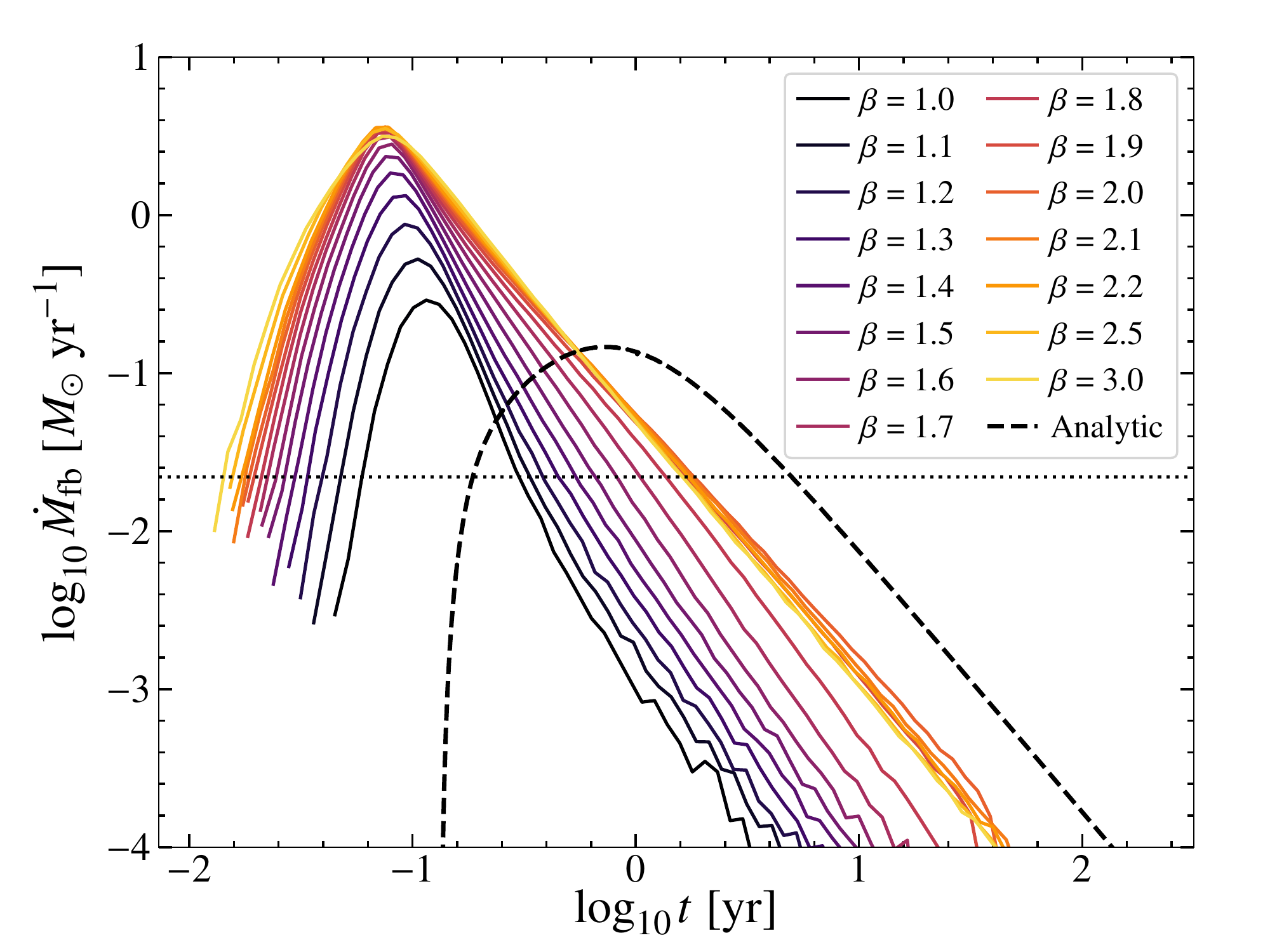}\\
\includegraphics[width=\columnwidth]{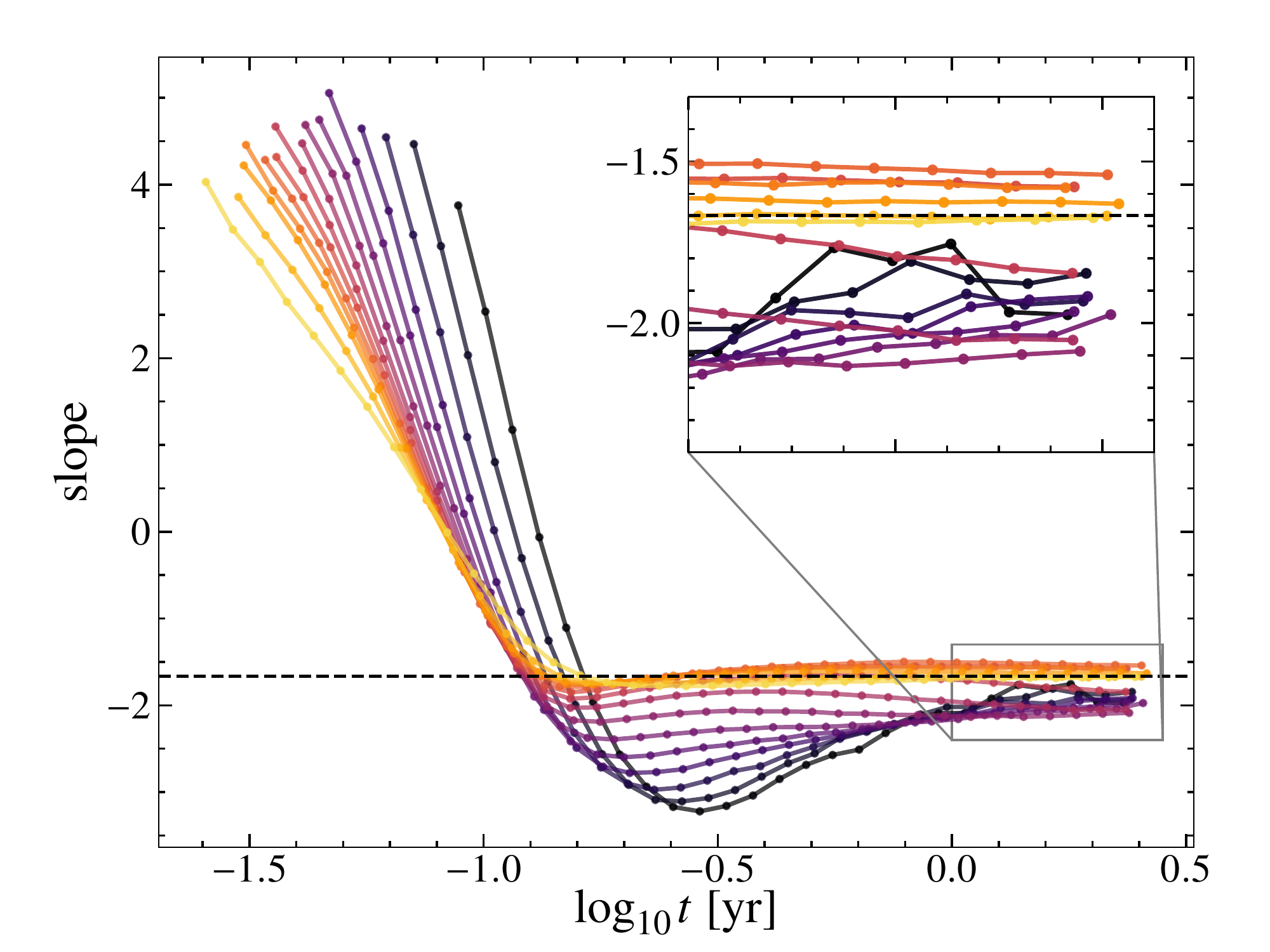}
\end{tabular}
\caption{\emph{Upper panel:} Fallback rates onto the SMBH as a
  function of time for the different orbits modelled.  As indicated in
  the legend, the penetration parameter increases from darker to
  lighter lines.  The black dashed line shows the rate computed
  assuming that the energy distribution is frozen in at the tidal
  radius, while the dotted horizontal line corresponds to the
  Eddington mass accretion rate for a $10^6\,{\rm M}_\odot$ SMBH.
  \emph{Lower panel:} Time evolution of the logarithmic derivative of
  each fallback rate presented in the upper panel.  The dashed
  horizontal line indicates the theoretical value of -5/3.
}
\label{fig:fallback}
\end{figure}

\begin{figure*}
\begin{tabular*}{\textwidth}{p{0.25\textwidth}@{}p{0.25\textwidth}@{}p{0.25\textwidth}@{}p{0.25\textwidth}}
\includegraphics[width=0.24\textwidth]{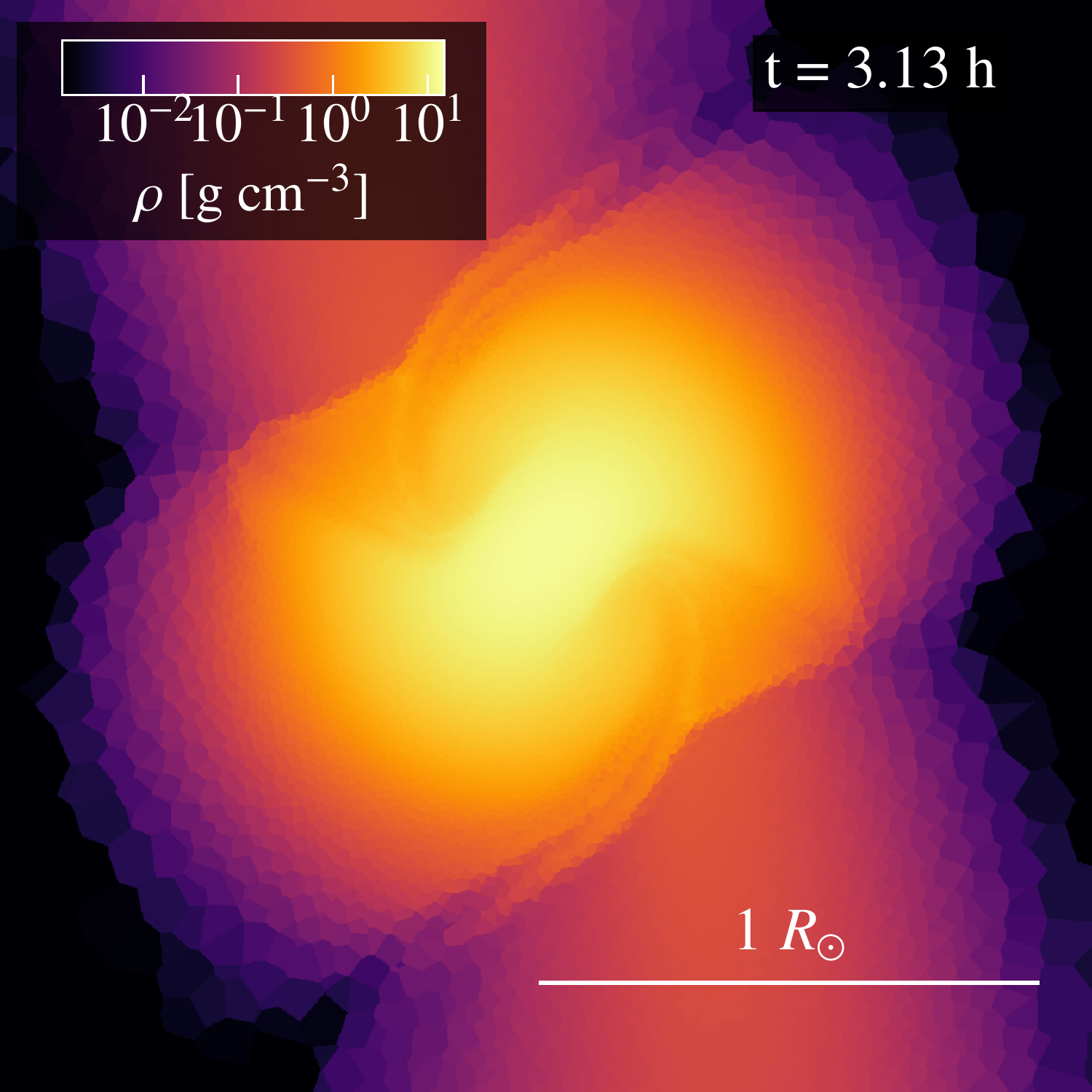} &
\includegraphics[width=0.24\textwidth]{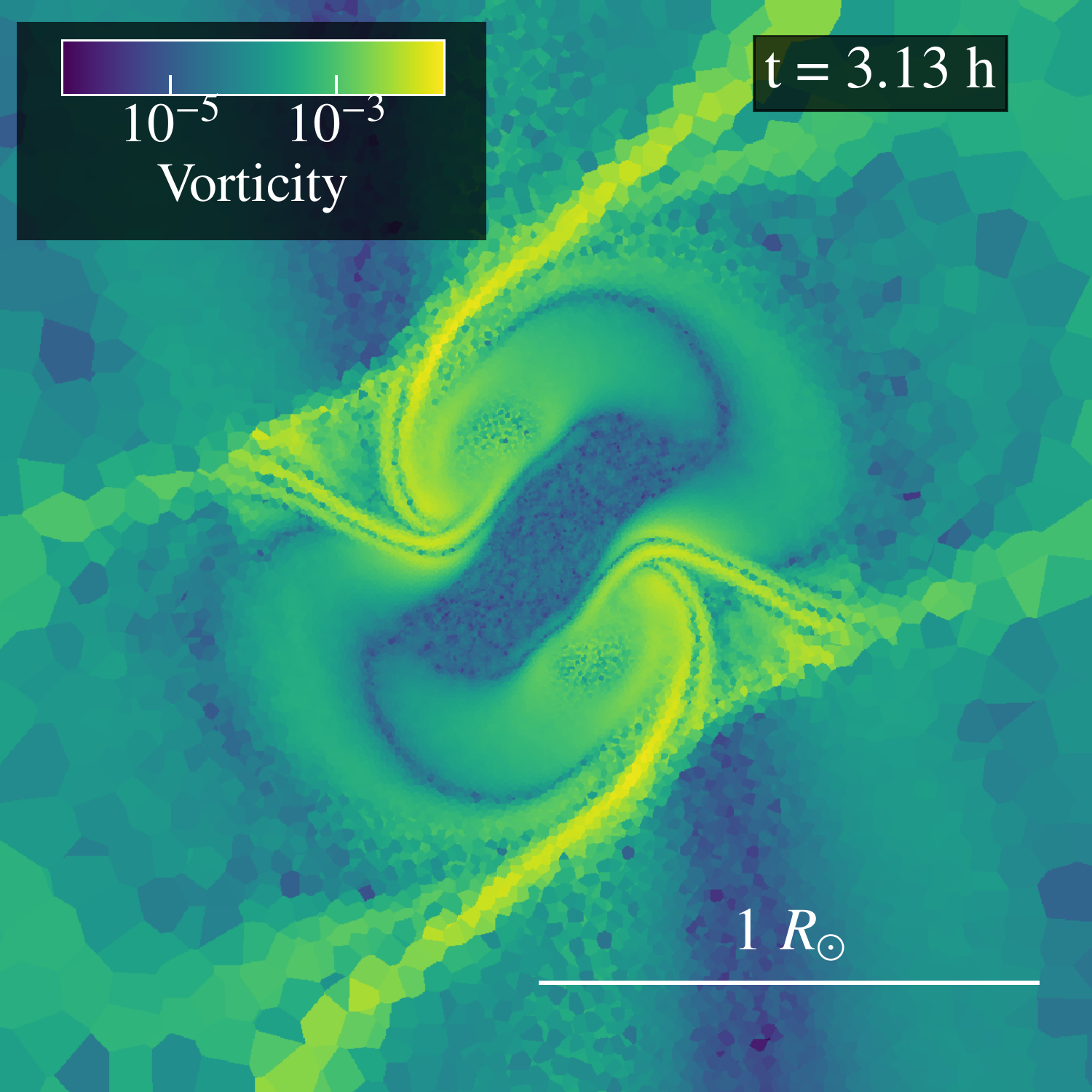} &
\includegraphics[width=0.24\textwidth]{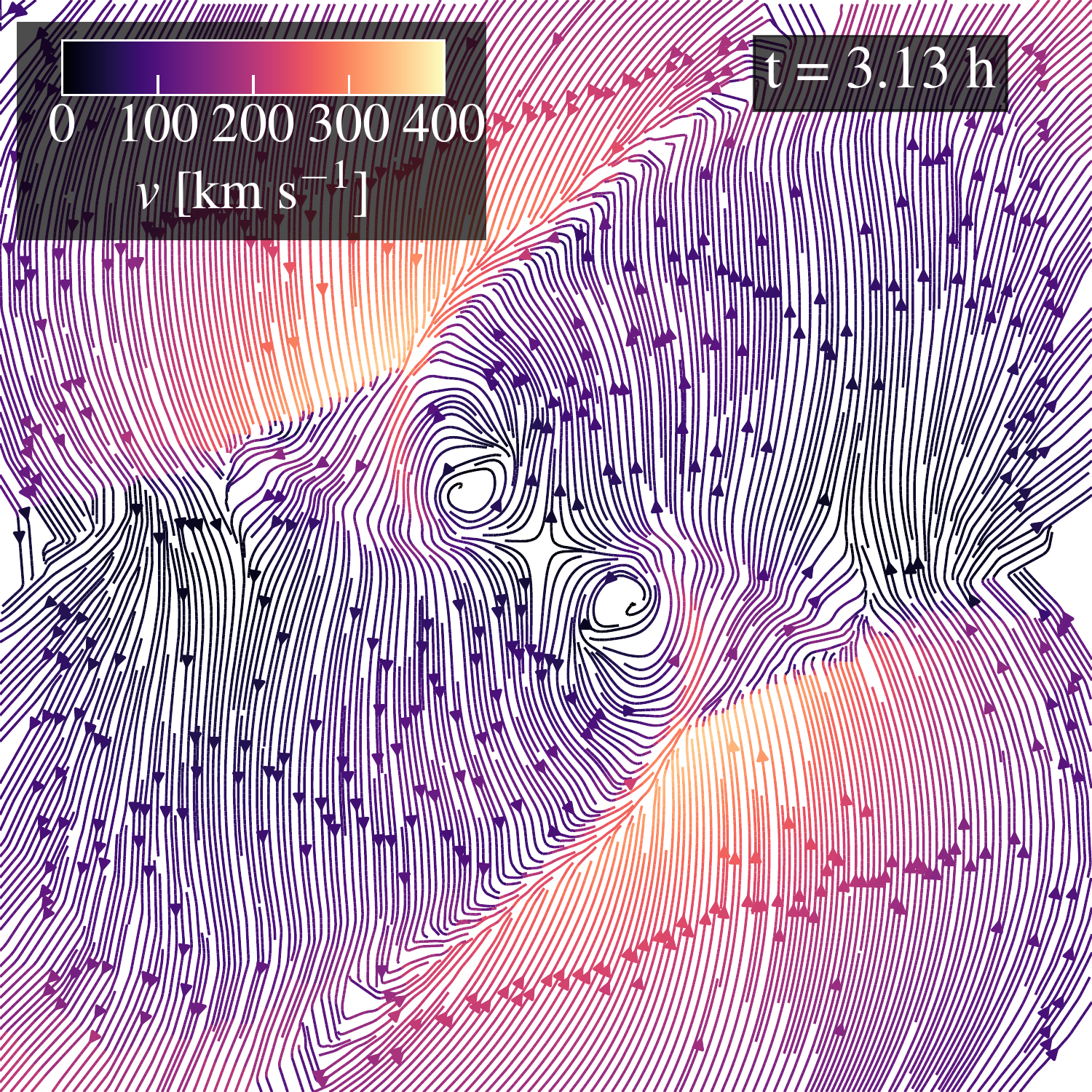} &
\includegraphics[width=0.24\textwidth]{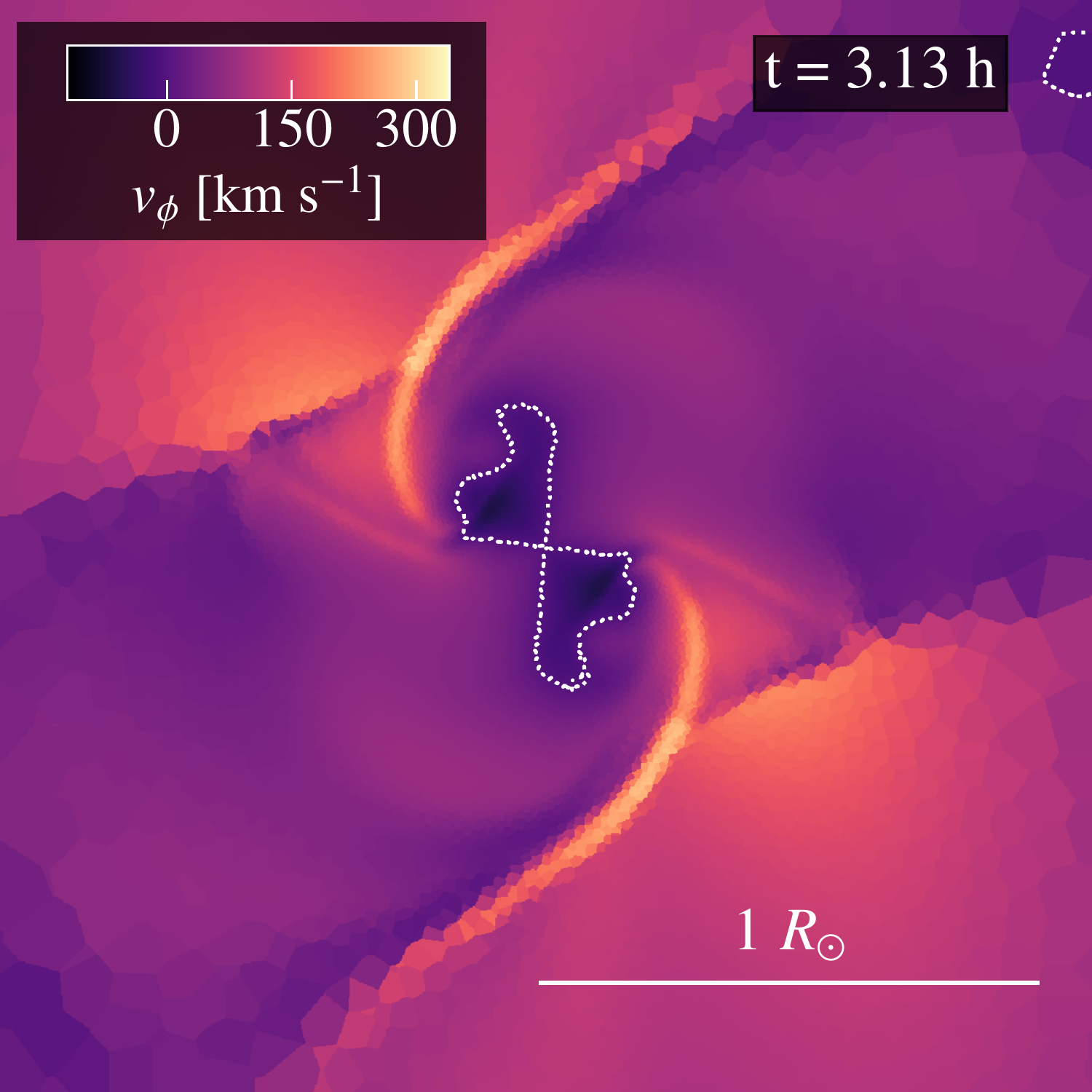} \\
\includegraphics[width=0.24\textwidth]{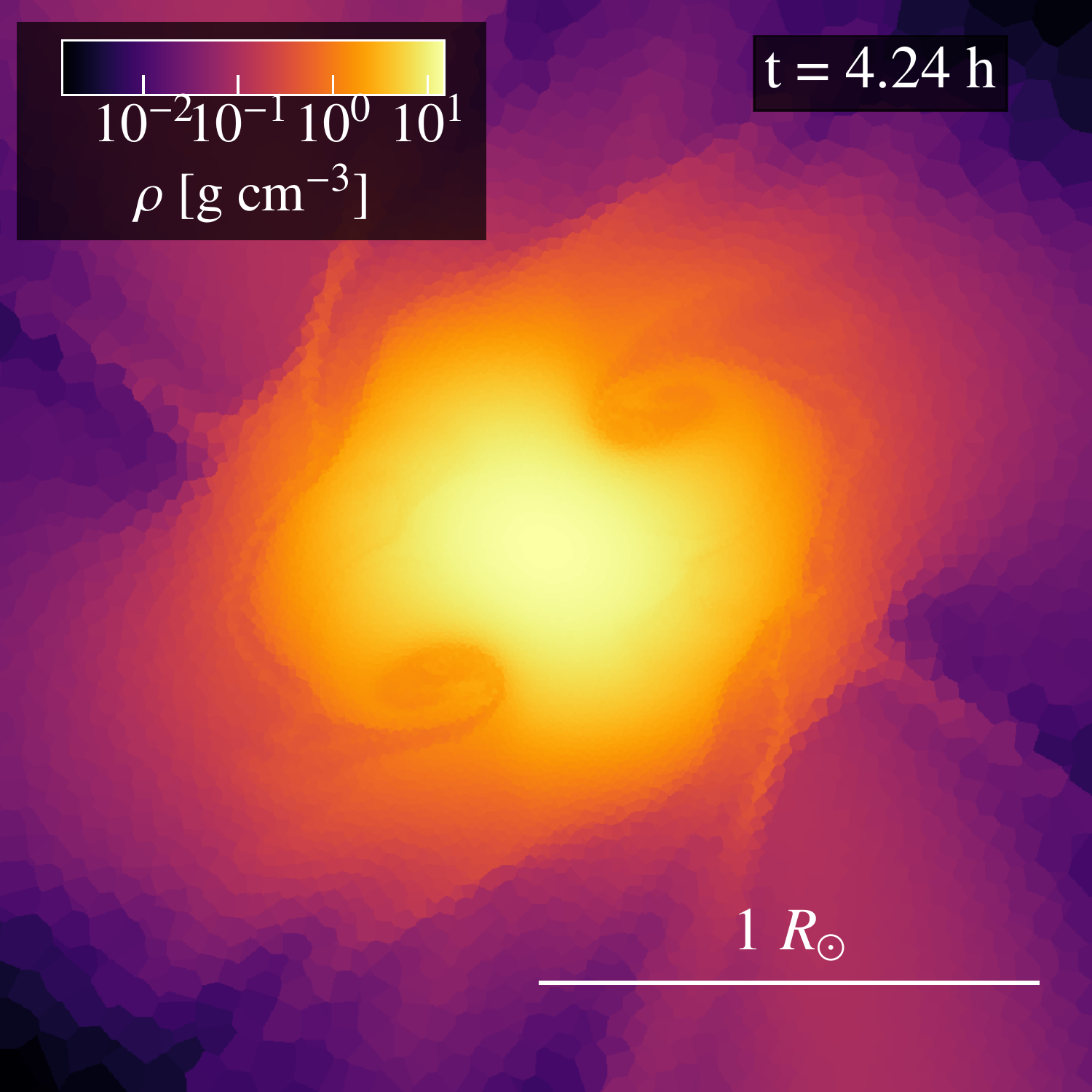} &
\includegraphics[width=0.24\textwidth]{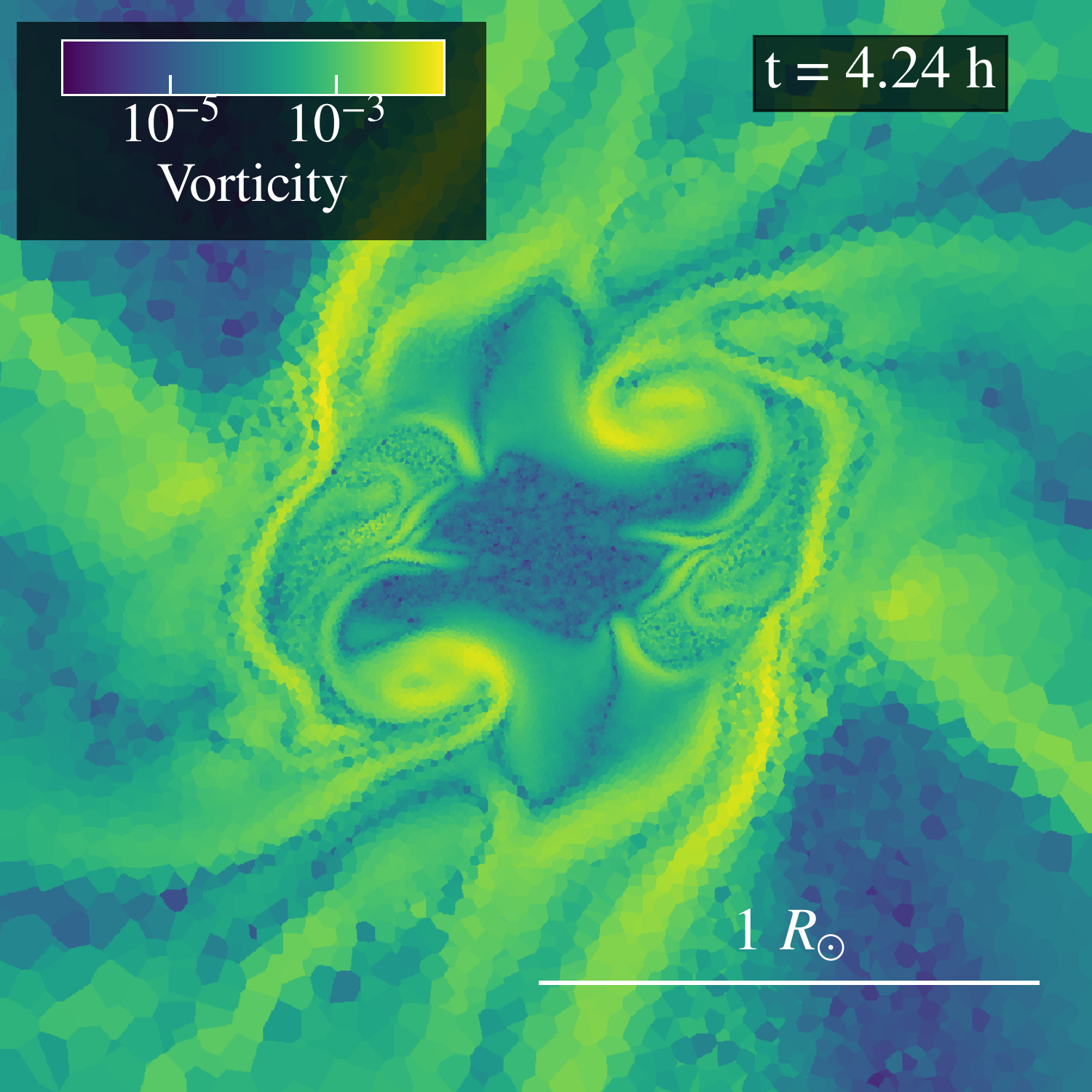} &
\includegraphics[width=0.24\textwidth]{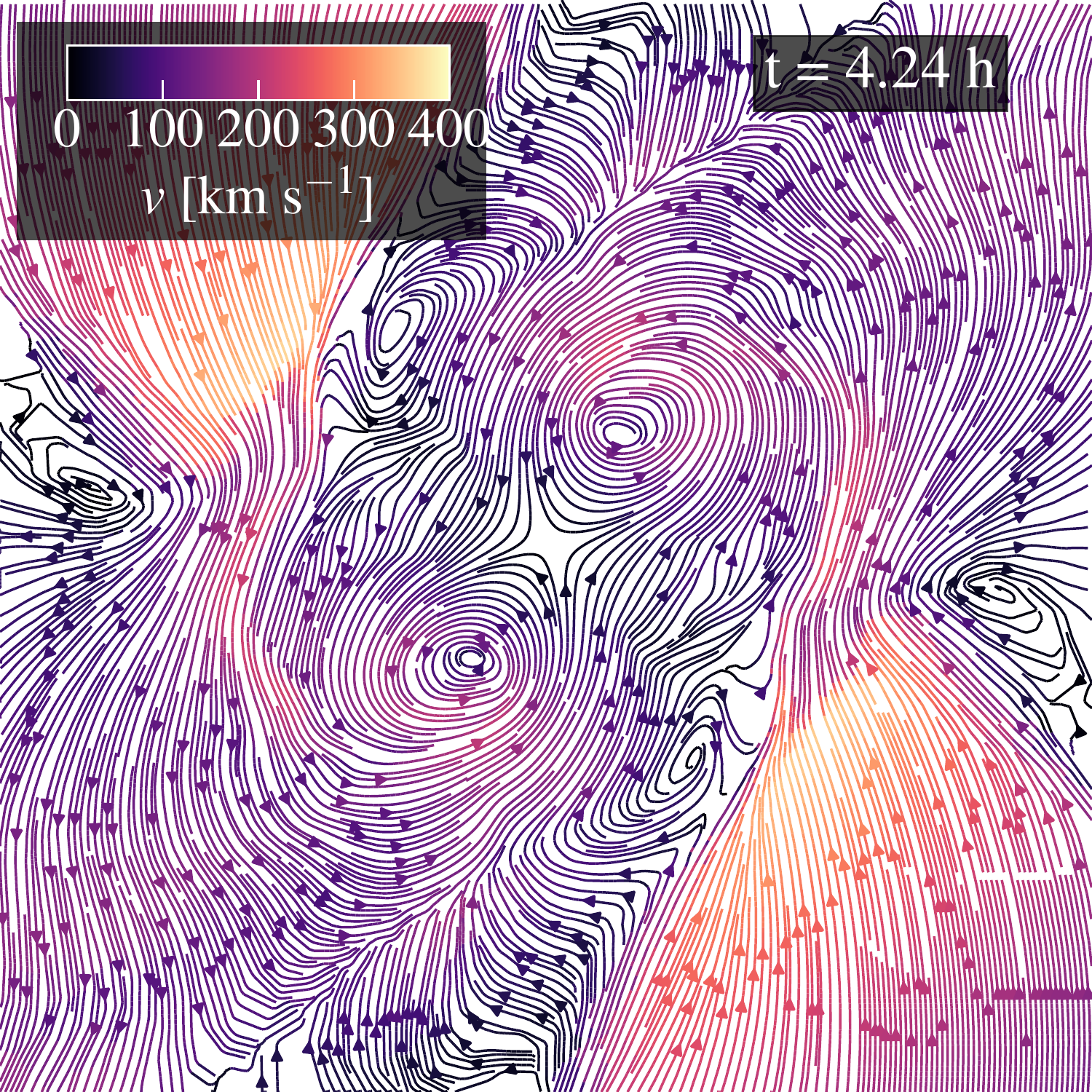} &
\includegraphics[width=0.24\textwidth]{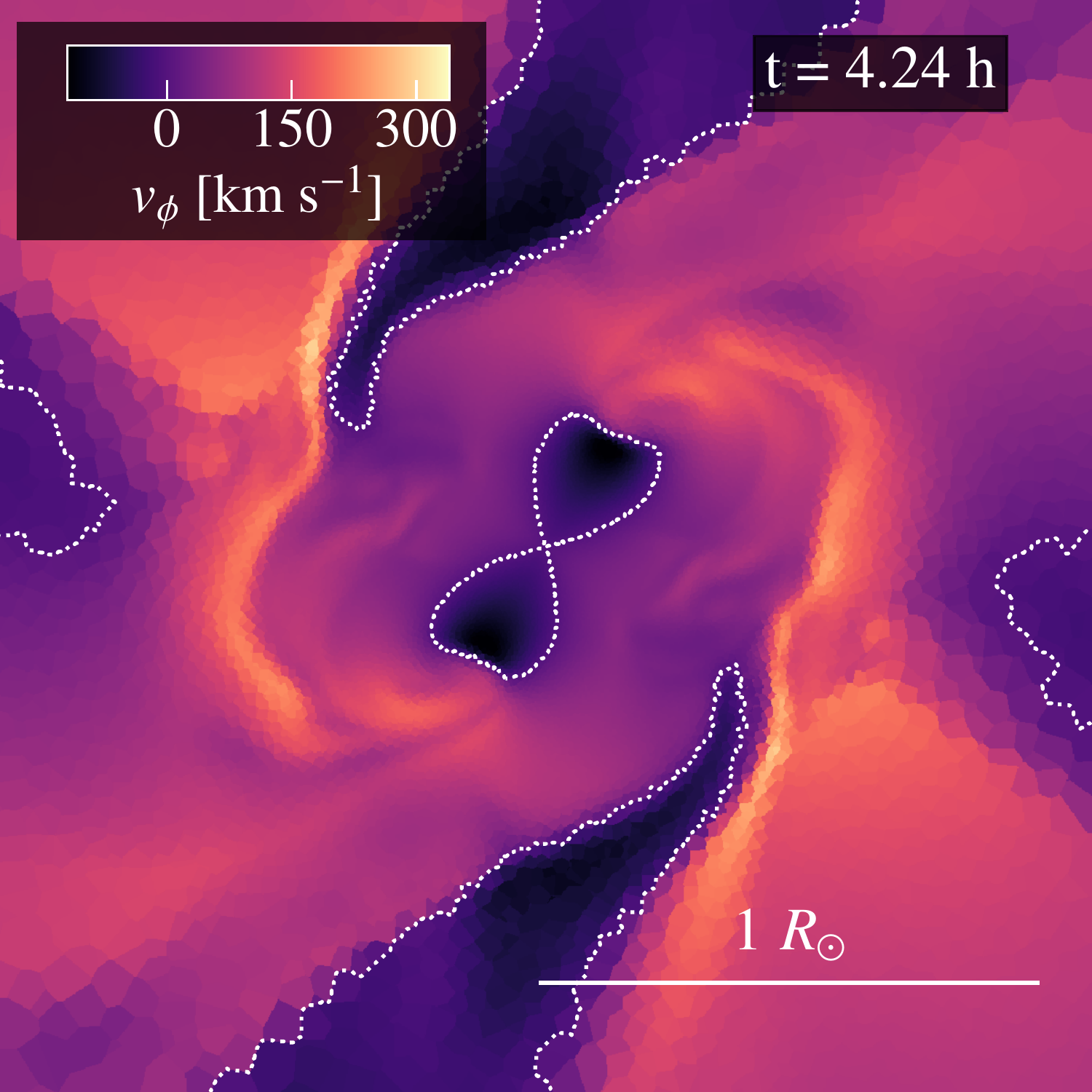} \\
\includegraphics[width=0.24\textwidth]{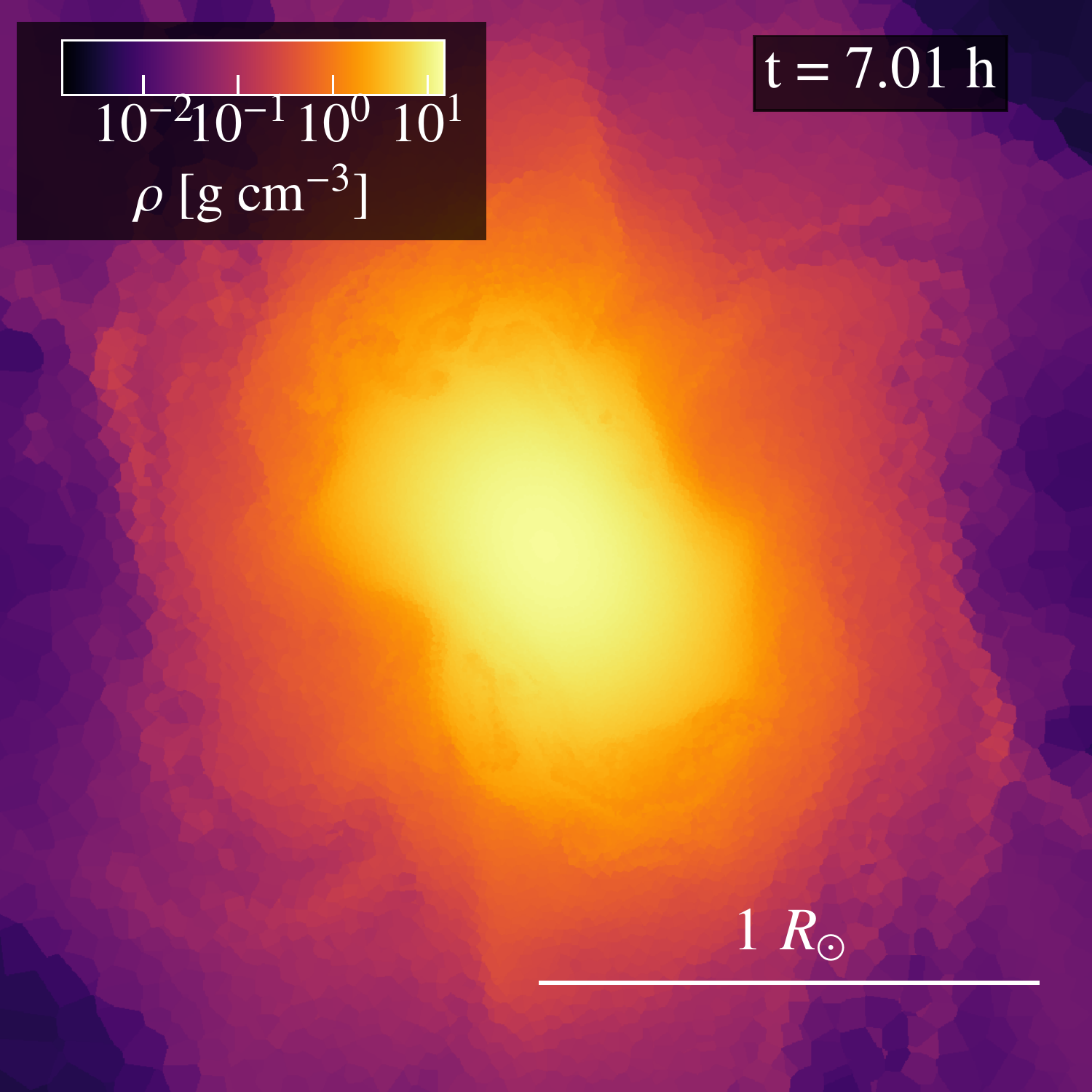} &
\includegraphics[width=0.24\textwidth]{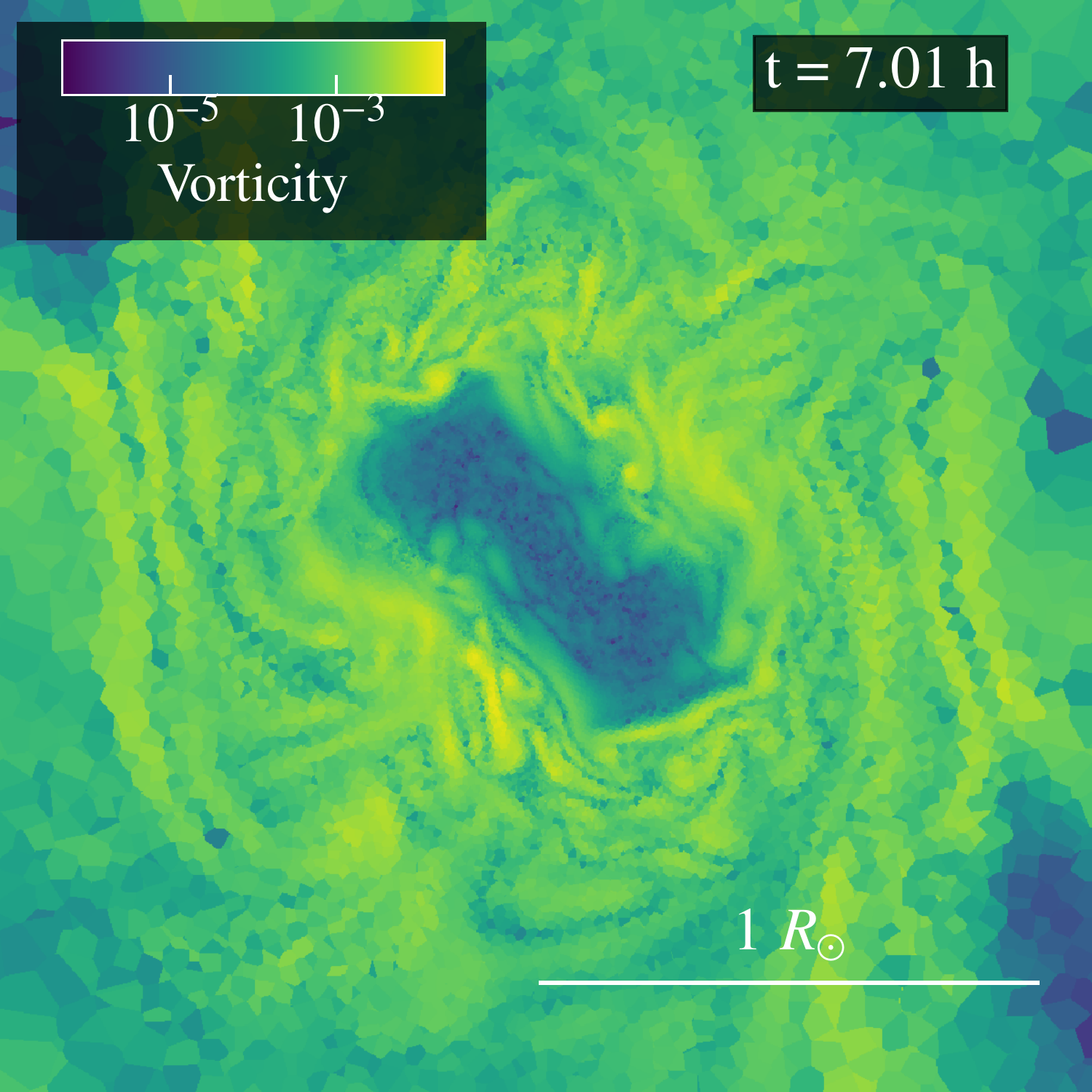} &
\includegraphics[width=0.24\textwidth]{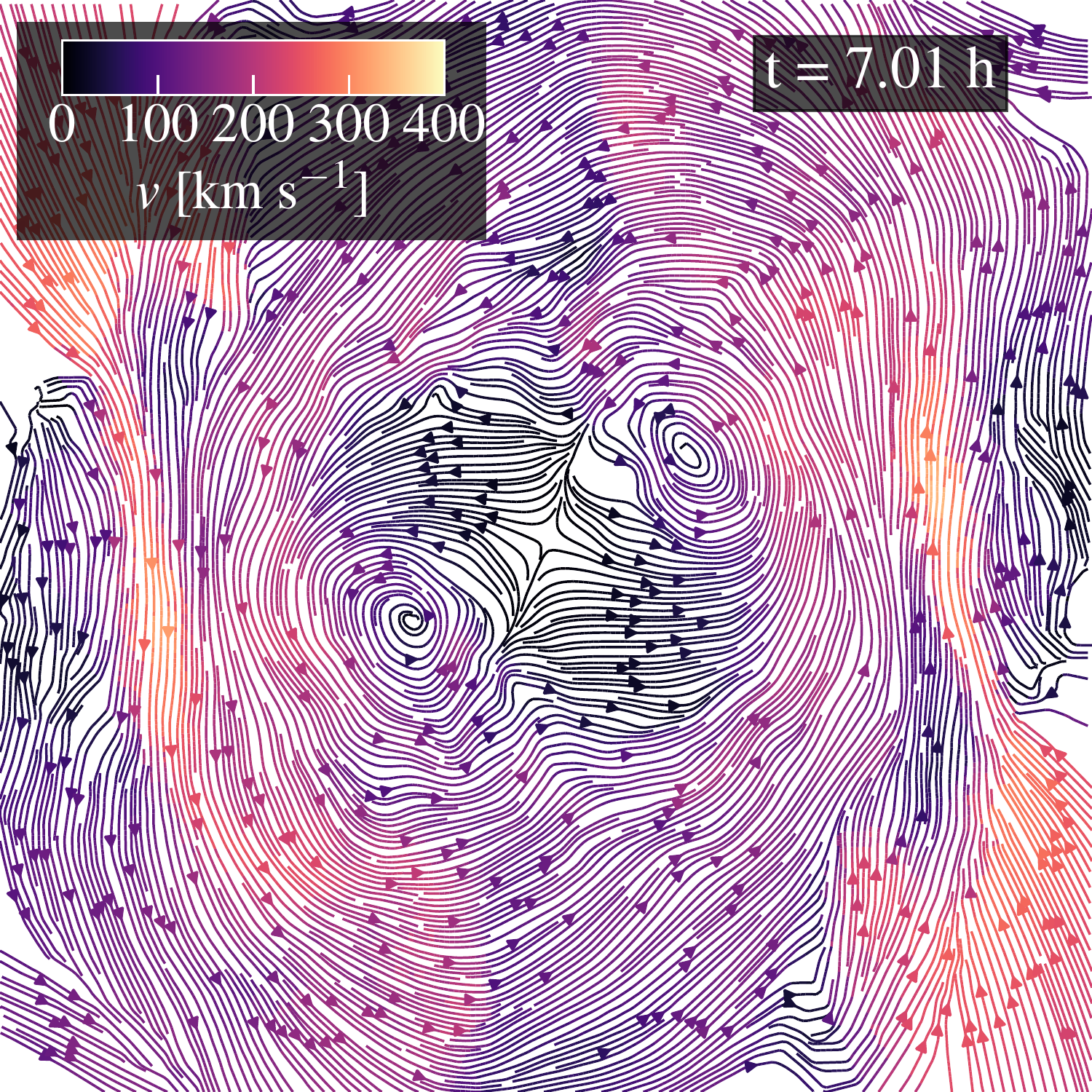} &
\includegraphics[width=0.24\textwidth]{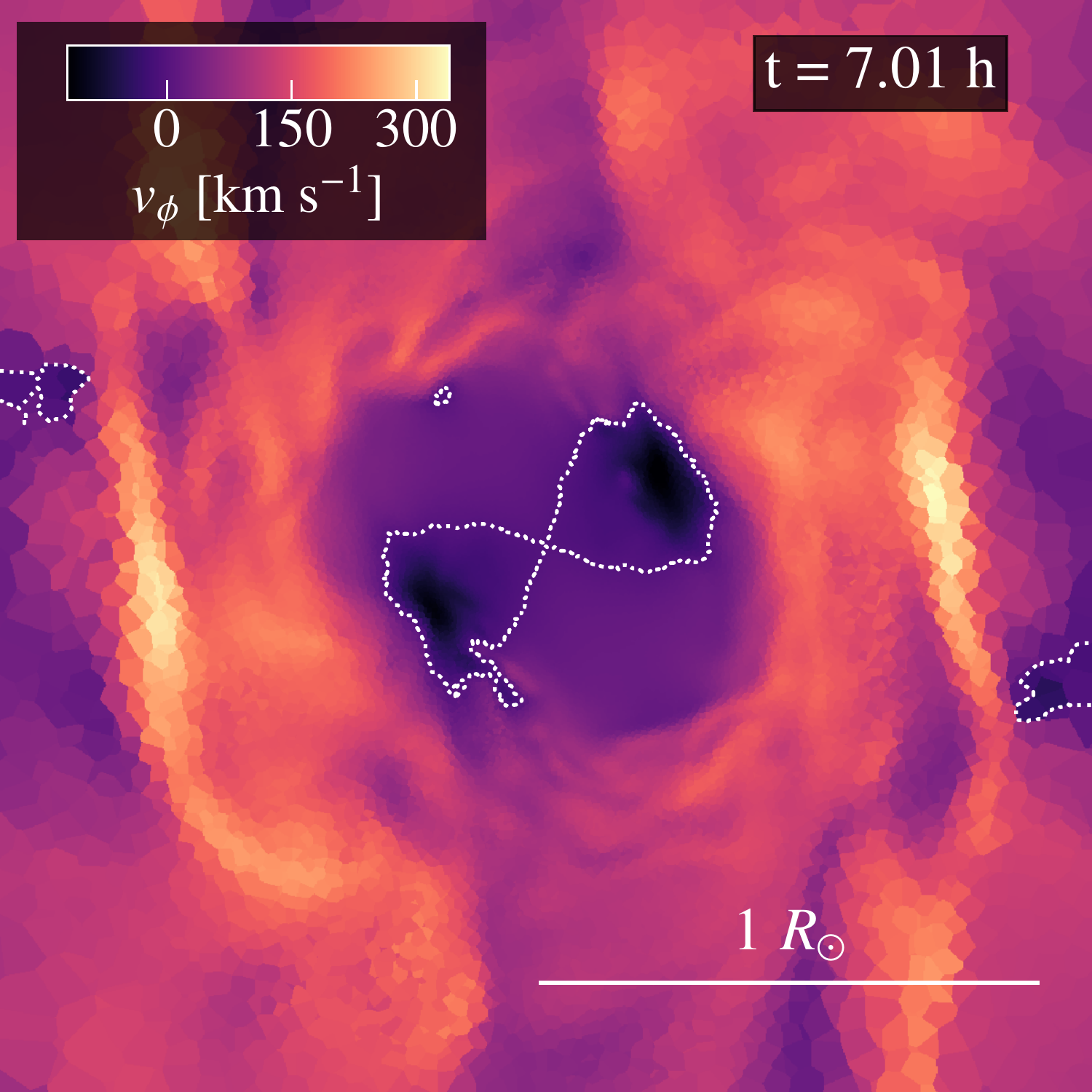} \\
\includegraphics[width=0.24\textwidth]{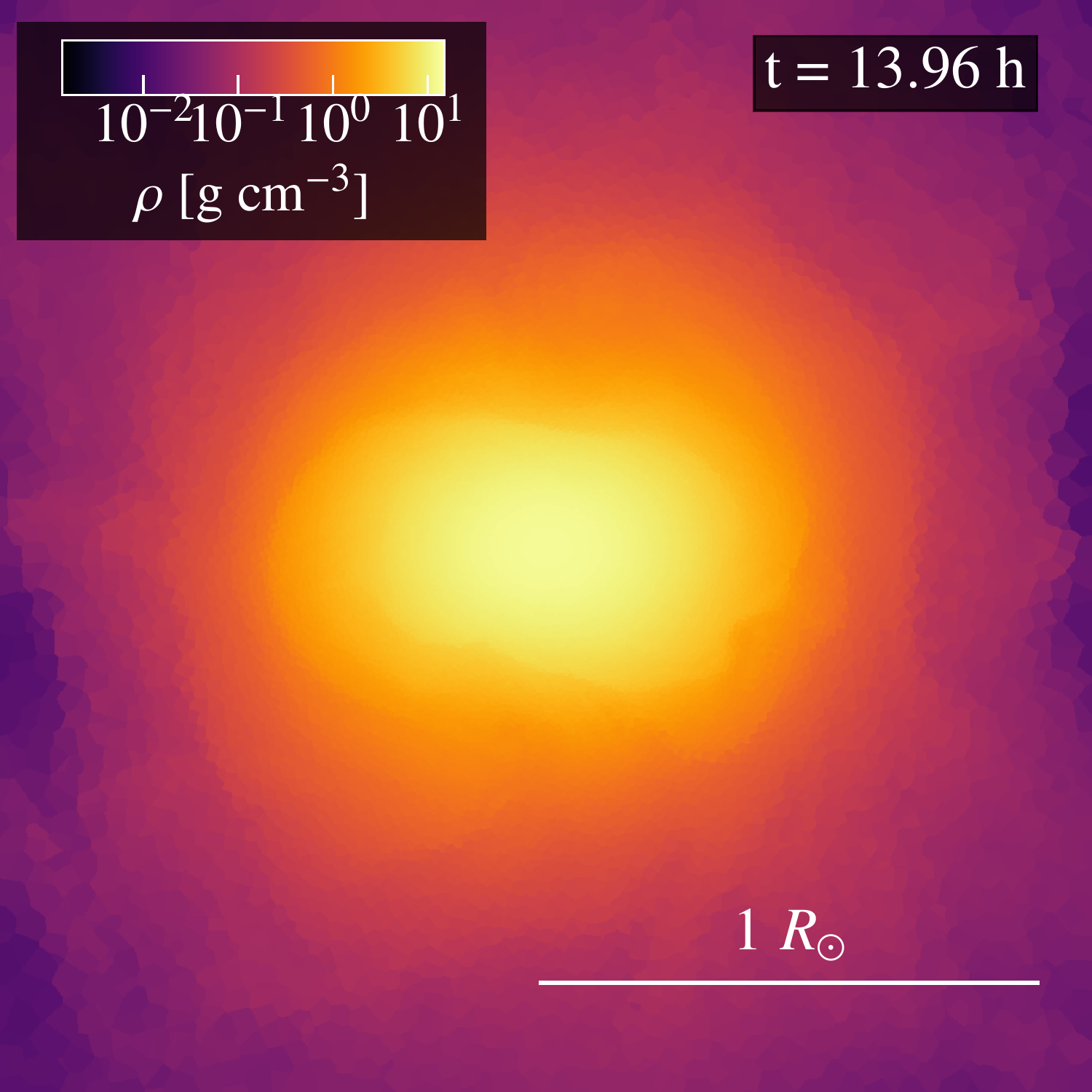} &
\includegraphics[width=0.24\textwidth]{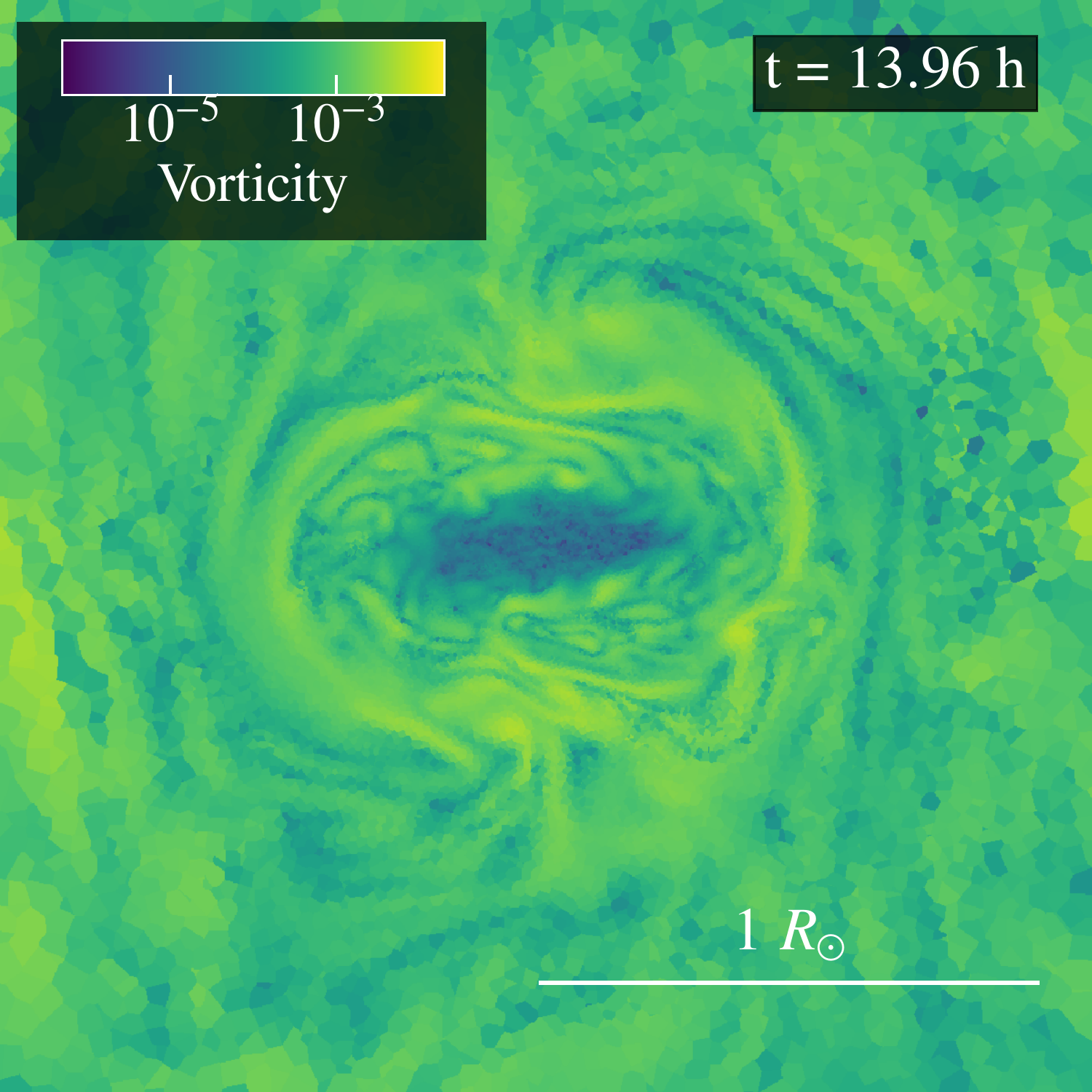} &
\includegraphics[width=0.24\textwidth]{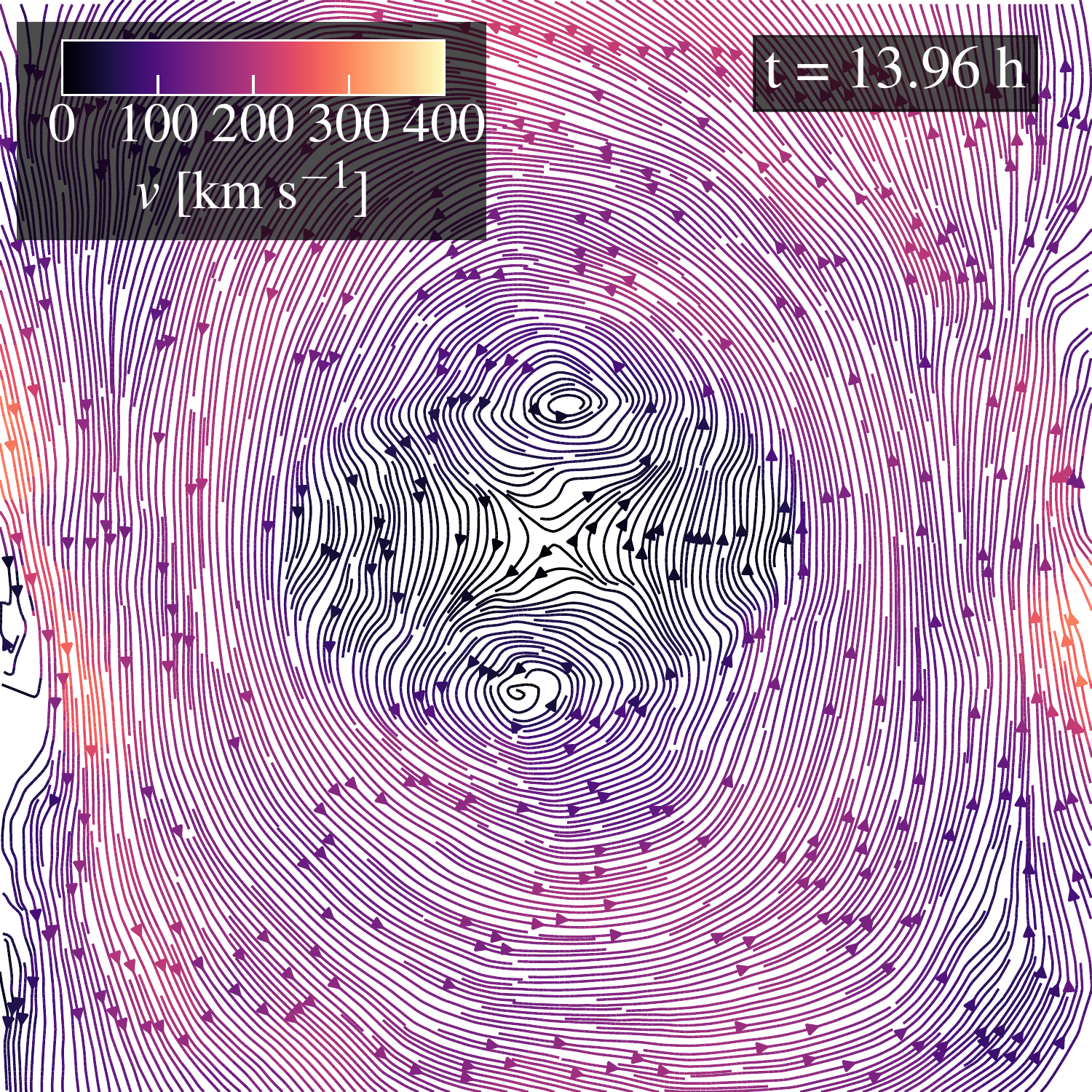} &
\includegraphics[width=0.24\textwidth]{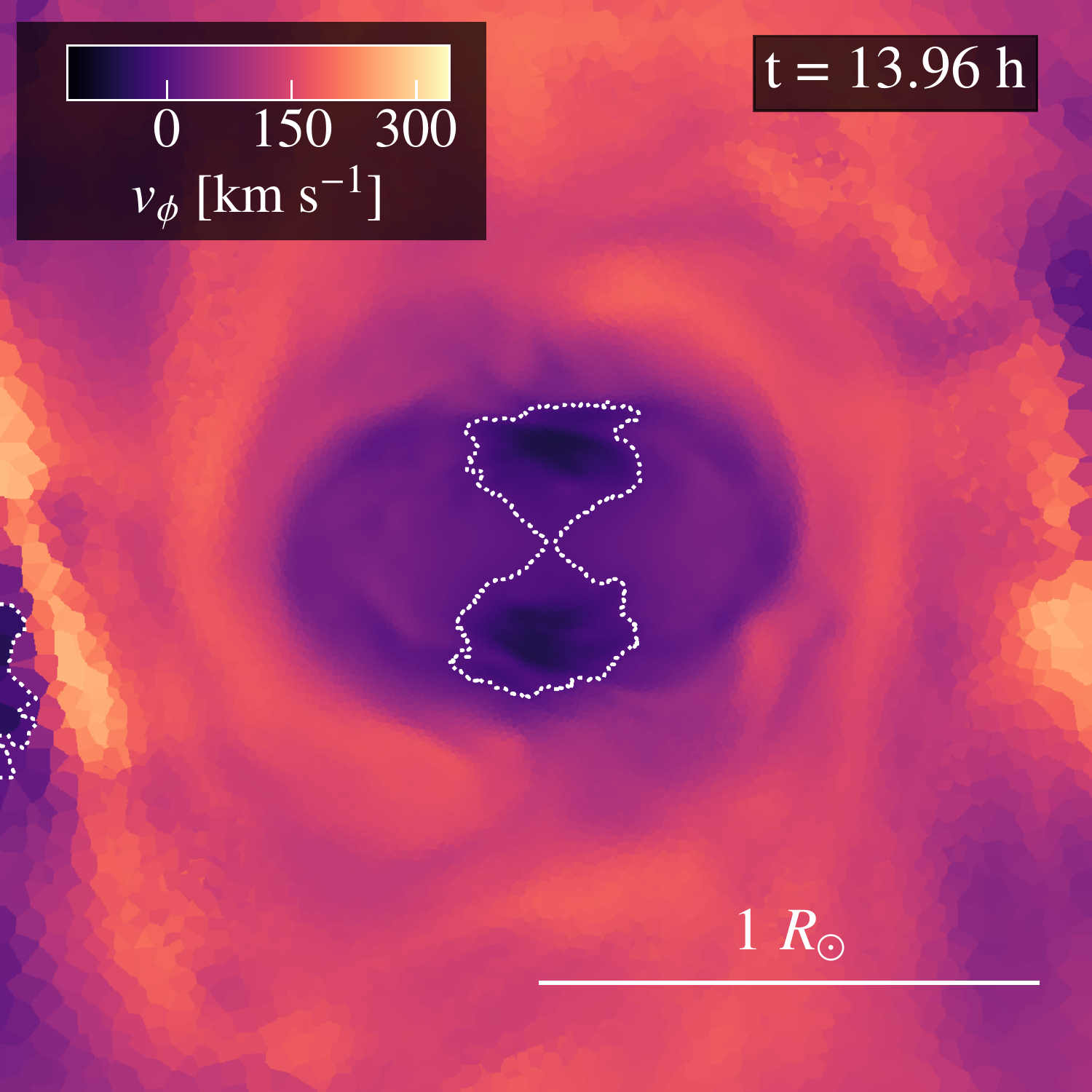} \\
\end{tabular*}
\caption{Time evolution of the surviving stellar core after a grazing
  encounter ($\beta=1.6$) with a SMBH.  From left to right, we show 2D
  slices of density, vorticity, total velocity and azimuthal velocity.
  All plots are centred on the core's centre of mass with a box side
  of $2\,R_\odot$, and the time label is such that $t=0$ corresponds to
  periapsis.
  {The dashed lines in the right-hand column mark the point where
  the azimuthal velocity is zero.}
  }
\label{fig:core_evolution}
\end{figure*}

\subsection{Energy distribution and fallback rate}

As shown by \citet{Rees1988}, during the disruption of a star on a
parabolic orbit roughly half of the material stripped from the star is
bound to the black hole. This gas will then fallback to the black
hole, powering a short period of activity.  The fallback rate of this
material as a function of time can be obtained using Kepler's third
law,
\begin{equation}
\dot M(t) = \dv{M}{E}\dv{E}{t} = \frac{(2\pi GM_{\rm BH})^{2/3}}{3}\dv{M}{E}t^{-5/3}.
\label{eq:fallback}
\end{equation}
The characteristic $t^{-5/3}$-decay
expected from these type of events \citep[see e.g.][]{Komossa2015}
is thus intimately related to the distribution of specific binding energy
($\dd{M}/\dd{E}$).

As we presented in the previous section, we have an
estimation of the total mass lost by the star, measured
at the end of our simulations.
We show the distribution of binding energy of this stripped material
for all impact parameters in Fig.~\ref{fig:energy}.
The dashed lines
in this figure show the expected spread in binding energy if the
distribution was `frozen in' at the tidal radius.  This approximation
is based on the consideration that inside the tidal radius only the black
hole gravity determines the energy distribution, and the internal
forces of the star are negligible.  Using this approximation, the
energy spread can be obtained by Taylor-expanding the
black hole gravitational potential across the star, yielding
\begin{equation}
\Delta E = \frac{GM_{\rm BH}}{r_t^2}R_*.
\end{equation}
The spread observed in our simulations is noticeably larger than this
estimate. This occurs because, once the star enters the tidal
radius, forces inside the star re-distribute some of the energy, and
the final distribution is no longer determined solely by the black
hole gravity.  For instance, \citet{Lodato2009} found that shocks in
the gas after the disruption promote the appearance of wings in the
tails of the energy distribution.
{Alternatively, \citet{Coughlin2016} argue that these
deviations result from the combination of the star's self-gravity
and the in-plane compression happening near periapsis.
}

Using equation~\eqref{eq:fallback} we can map the distribution of
binding energy to the fallback rate of gas onto the black hole, which
is shown in Fig.~\ref{fig:fallback} for every impact parameter.  It is
important to stress that this fallback rate does not translate
directly to an accretion rate onto the black hole, since the gas would
likely first settle into an accretion disc, where the dissipation of
energy and angular momentum occurs on a viscous timescale.

Under the `frozen in' approximation, the energy distribution is
completely determined by the fraction of stellar mass at each slice
pointing towards the black hole.  We now use the formalism presented
by \citet{Lodato2009} to compute the energy distribution and
corresponding fallback rate, which can be expressed as
\begin{equation}
\dv{M}{x} = 2\pi\int_0^{H_x} \rho_* h \dd{h},
\label{eq:mass_dist}
\end{equation}
where $x$ is the radial coordinate inside the star,
$H_x = \sqrt{R_*^2-x^2}$ is the radius of each slice, and $\rho_*$ is
the one-dimensional density profile generated by MESA.  Because the
latter is not an analytical profile, we numerically integrate
equation~\eqref{eq:mass_dist} to obtain the black dashed line in
Fig.~\ref{fig:fallback}.  Despite the crudeness of assuming that the
energy distribution is frozen in at the tidal radius, it is still frequently
being used as an approximation to compute fallback rates coming from
the stellar disruption \citep[e.g.][]{GallegosGarcia2018}, yet the
accuracy of this approach is quite limited based on our results.

In the lower panel of Fig.~\ref{fig:fallback} we show the time
evolution of the logarithmic derivative of the fallback rates,
i.e.~the slope.  From the inset in this plot it is clear that for the
encounters that result in a partial disruption ($\beta\lesssim$1.8)
the slope approaches steeper values than the theoretical expectation
at late times.  As explained by \GRR, this is due to the gravitational
influence of the surviving core, countering the black hole's tidal
force for the closest gas.
This influence {monotonically} depletes gas at 
{lower energies after the peak} (see Fig.~\ref{fig:energy},
 lower panel), which is the
material that determines the asymptotic value of the fallback
rate.
On the other hand, we find that the cases $\beta=1.9, 2.0, 2.1, 2.2$
present shallower slopes compared to $-5/3$.
This behaviour is also found by \GRR, and comes from the fact that in the
borderline cases where the star is barely completely destroyed, some
of the core material slowly {shrinks} as it is pulled apart.
{Consequently, and in contrast with more grazing encounters,
these cases present an increase of gas towards lower energies
(see Fig.~\ref{fig:energy}, upper panel).}

For deeper encounters ($\beta\gtrsim2.5$), the star is quickly destroyed at
pericentre, and thus the slope is consistent with $-5/3$.
{This occurs because the energy distributions do flatten at low energies.}
{However, it is important to clarify that the large dip observed around 
$E=0$ is the result of our iterative procedure to determine the stripped mass. 
As previously discussed, because the tidal forces vanish towards the 
debris' centre of mass, this procedure always yields a non-zero amount of 
self-bound gas, which  is thus excluded from the determination of the 
energy distribution.
However, this affects only material with fallback timescales of over a 
decade, and does not change the rates shown in Fig.~\ref{fig:fallback}.
In any case, because $\mathcal{F}\sim 1$ at all times for total 
disruptions (see equation~\ref{eq:f}), the gas considered as self-bound
by this method would eventually become negligible 
if we were to run the simulation for a long enough time.}

\begin{figure}
\centering
\includegraphics[width=\columnwidth]{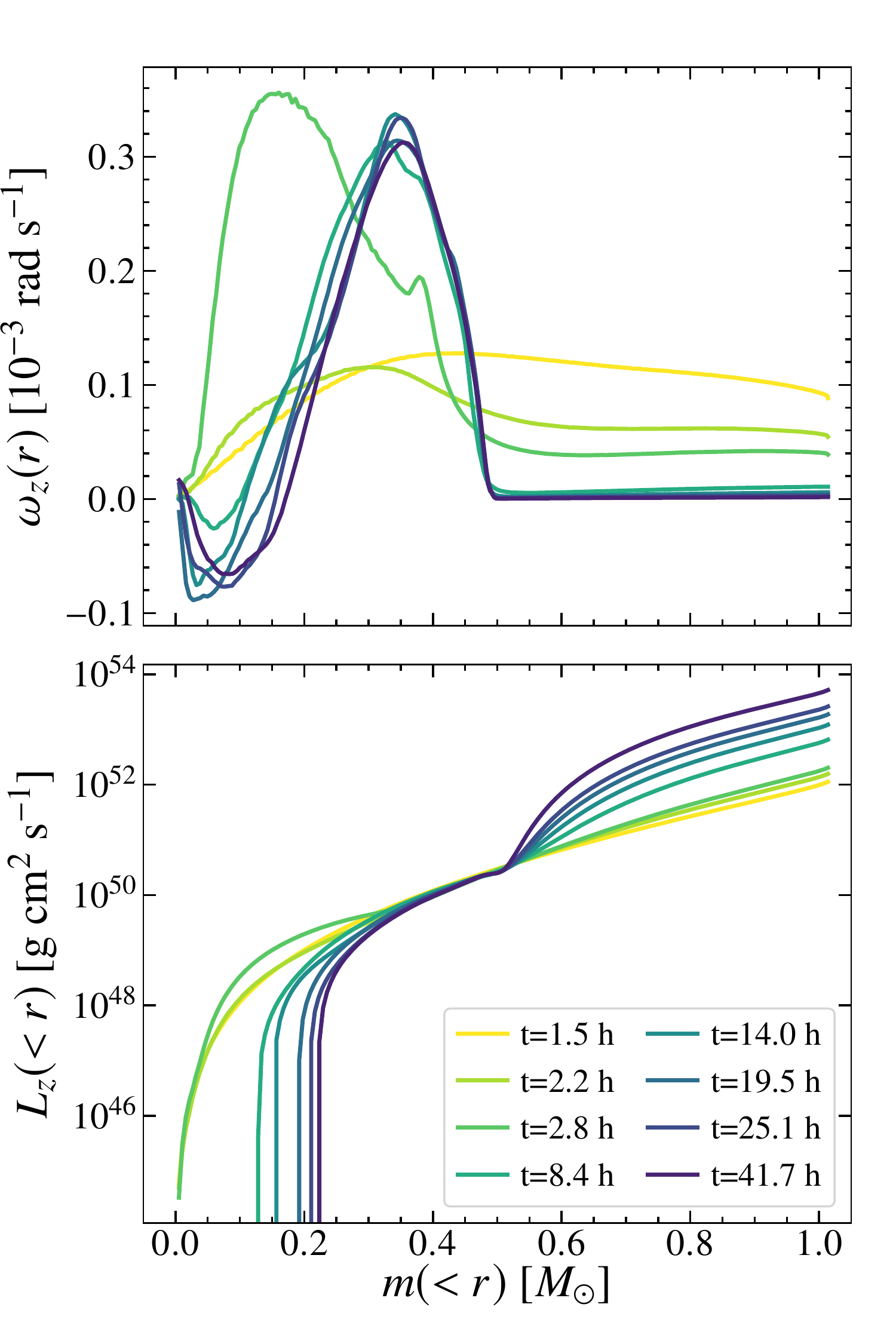}
\caption{\emph{Upper panel:} Time evolution of the surviving core's
  angular frequency during the interaction.  \emph{Lower panel:} Time
  evolution of the star's angular momentum profile.  In both panels,
  time goes from lighter to darker lines. Notice that we have replaced
  the radial coordinate by the enclosed mass, since the extent of the
  gas changes dramatically during the interaction.}
\label{fig:omega}
\end{figure}

\section{Hydrodynamics of the surviving core}
\label{sec:core}

As the star approaches the black hole, it is stretched into a prolate
spheroid along the direction of motion. Once the stars reaches
periapsis, its leading edge is slightly closer to the SMBH, where the
tidal forces produce a torque on the whole star, effectively inducing
some level of rotation \citep[see e.g.][]{Guillochon2009}.
Consequently, following a partial disruption, the surviving stellar
core rotates as it breaks away from the black hole.  The complex fluid
motions produced during these later stages can be studied with high
accuracy using \arepo, thanks to its quasi-Lagrangian nature and high
hydrodynamical accuracy compared to particle-based techniques.

To study the hydrodynamical evolution of the surviving core after a
grazing encounter we choose $\beta=1.6$ as a representative case.  In
this example, roughly half of the star's mass remains bound to the
core (see Fig.~\ref{fig:deltam}), although most of our conclusions
apply also to other instances where the stellar core survives.  The
early evolution of the star after periapsis is shown in
Fig.~\ref{fig:core_evolution} with slices in the $x-y$ plane.  The
density slices (left column) show the material as it is being stripped
from the star, with some of the bound material forming a diffuse halo
around the core.  To capture the complexity of the motions, we analyse
the vorticity of the fluid, defined as
\begin{equation}
w = \norm{\curl{\vb{v}}}.
\end{equation}
We show slices of this quantity inside the star in the second column
of Fig.~\ref{fig:core_evolution}.  In contrast to the density field,
the vorticity of the fluid shows plenty of substructure within the
surviving core, mainly in the form of small filaments.  This points to
a turbulent evolution of the star after the encounter.  Furthermore,
the substructures observed seem to indicate that the material inside
the core is stirred up by this turbulence, probably inducing
significant mixing in the gas.

\begin{figure*}
\begin{tabular}{ccl}
\includegraphics[width=0.3\textwidth]{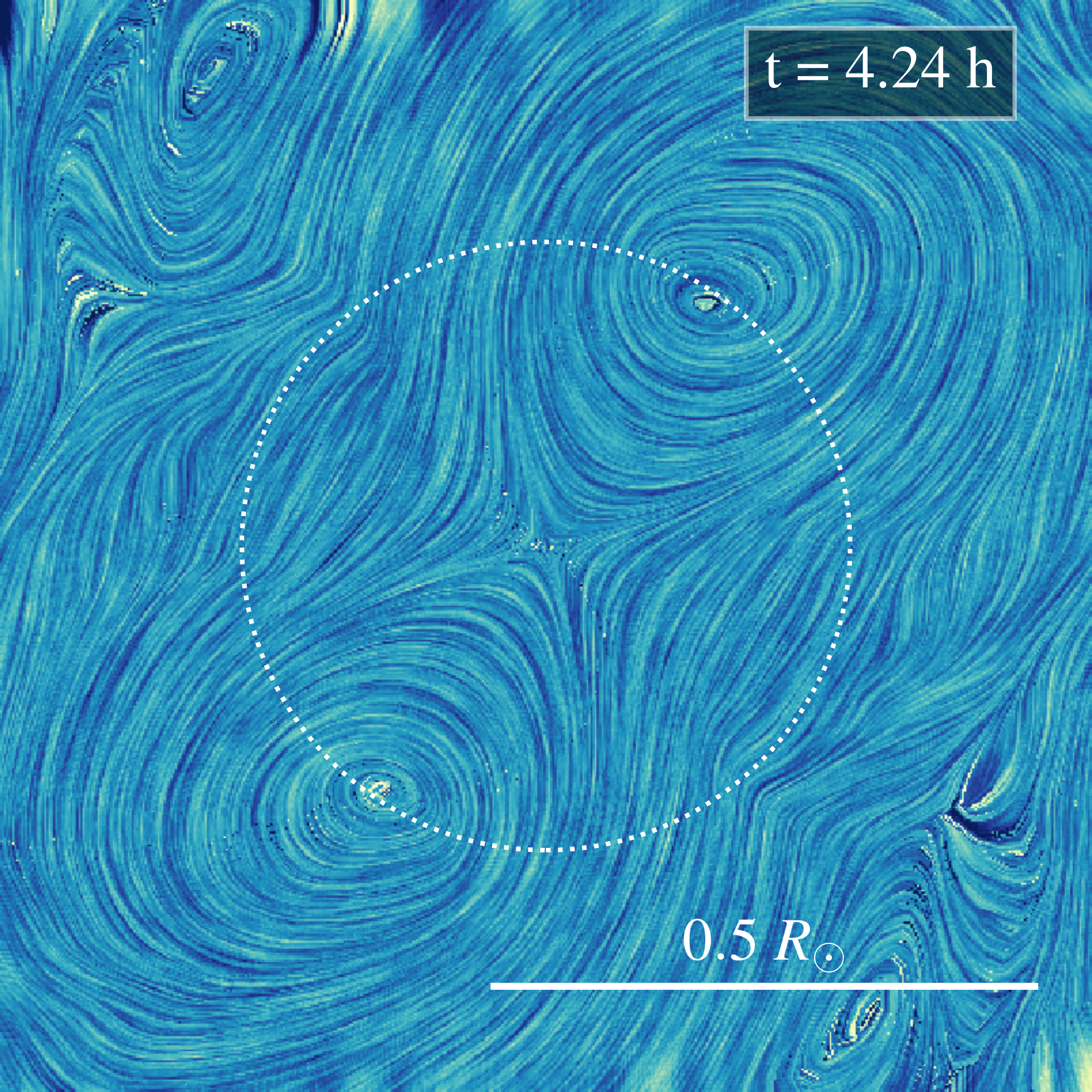} &
\includegraphics[width=0.3\textwidth]{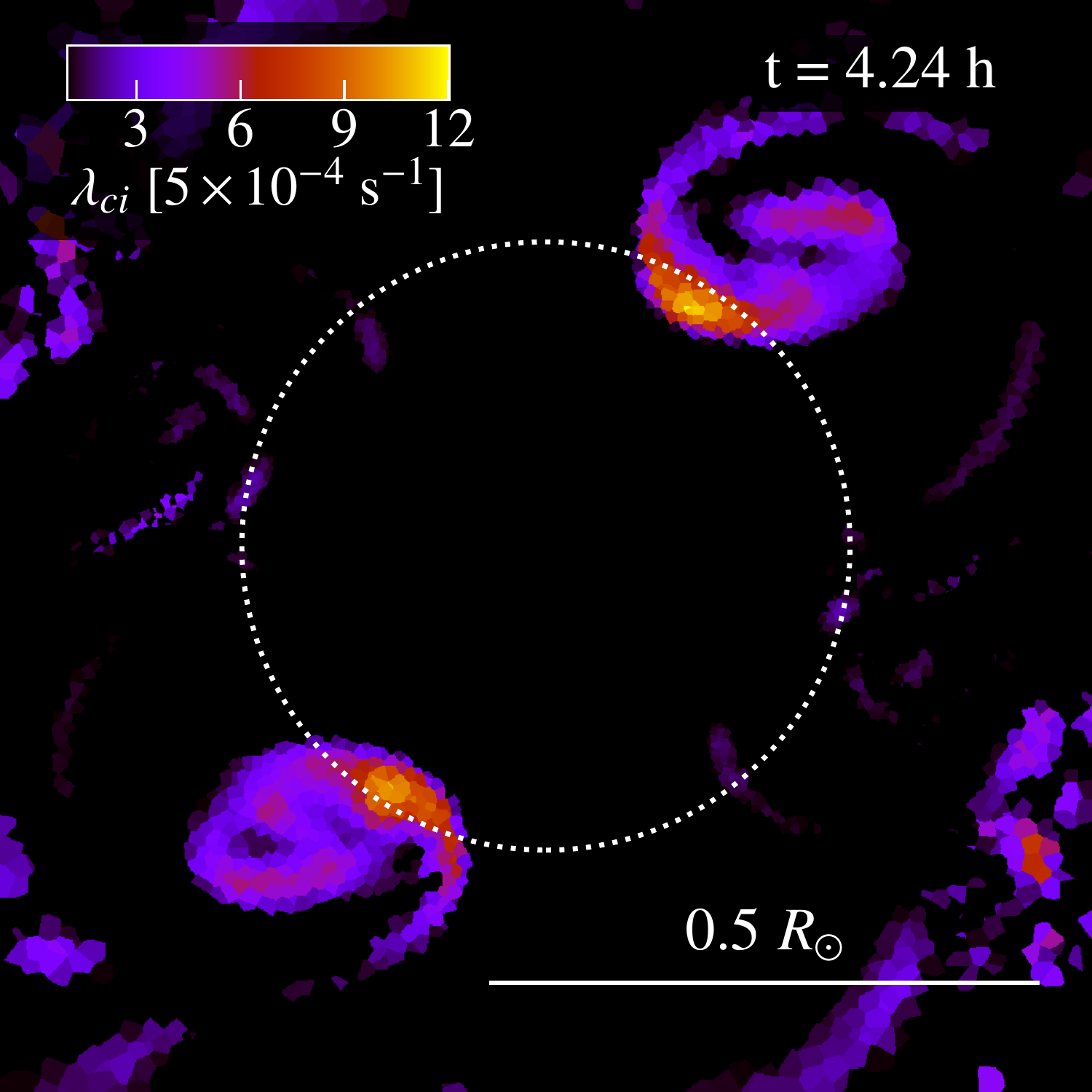} &
\includegraphics[width=0.33\textwidth]{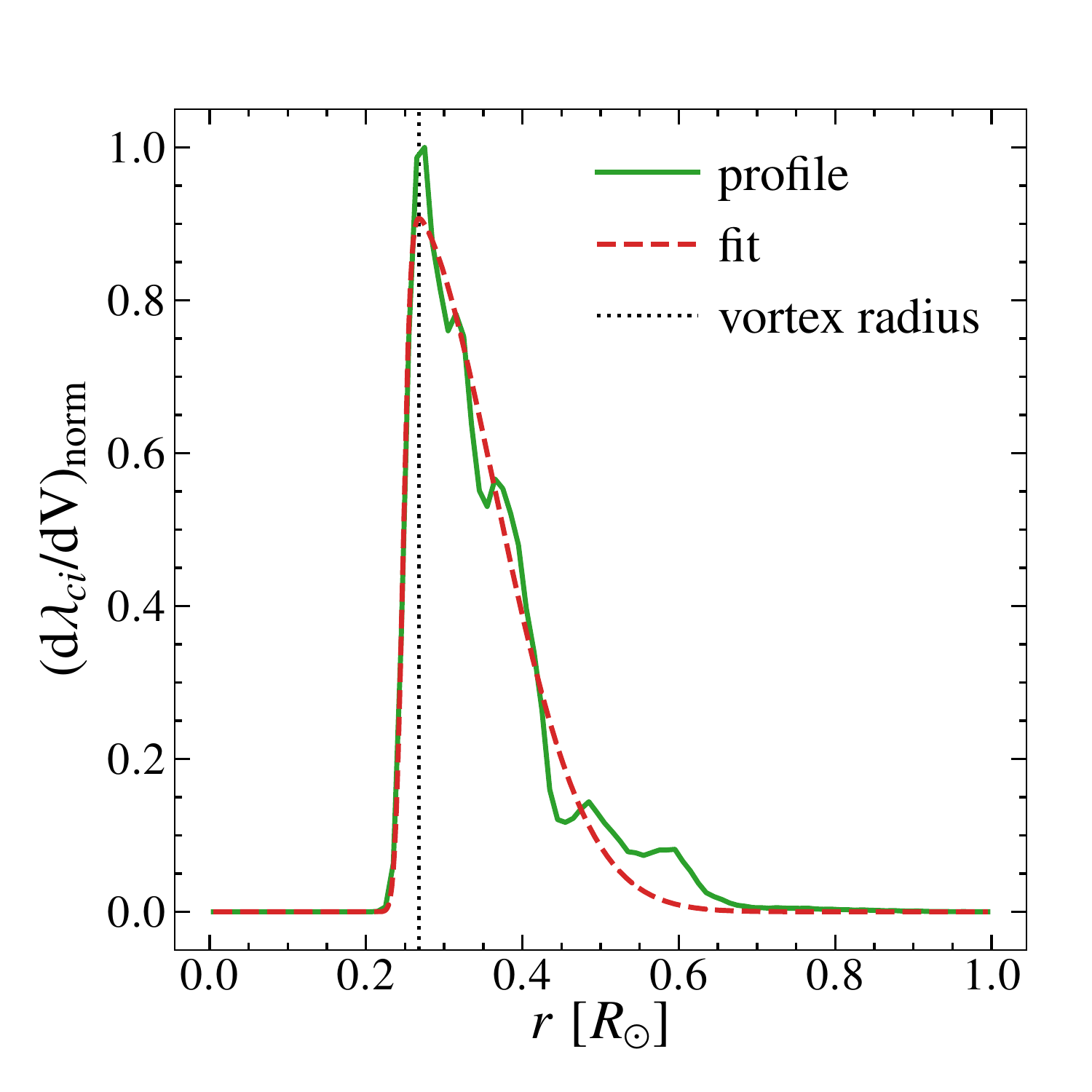}
\end{tabular}
\caption{Example of the prominent vortices developed inside the
  surviving core.  The left-hand panel shows the morphology of the
  velocity field inside the core using the line integral convolution
  method, while the center panel displays a slice of the ``swirling
  strength'' $\lambda_{ci}$ of the same gas.  The right-hand panel
  shows the density profile of $\lambda_{ci}$ {(normalized to the
  maximum value)}, where the peak
  indicates the radial position of the vortex pair, as shown in each
  panel with a dotted line.  }
\label{fig:vortices}
\end{figure*}

These complex motions are induced by the differences in velocity with
respect to the material surrounding the star and the gas that is being
stripped, which can be observed in the two right columns of
Fig.~\ref{fig:core_evolution}. The total velocity in the reference
frame of the star (middle right column), displayed with stream lines,
clearly shows the presence of two prominent vortices on opposite sides
of the core, formed as part of the re-accretion of some initially
stripped material.  This vortex pair rotates with the surviving core,
although the gas trapped inside has much lower azimuthal velocity
(Fig.~\ref{fig:core_evolution}, right column) than the rest of the
surrounding gas, in fact even reaching negative (counterrotating)
values. Hence the star is constantly being stirred up by these
movements, driving the turbulent behaviour. We note that the vortices
are still present by the end of our simulation, roughly after 50
stellar dynamical times, which shows the high persistence of this
structure.

Motivated by these complex motions, we study the evolution of the
stellar rotation during the interaction.  Since the star's movement
occurs in the $x-y$ plane, the relevant axis is the $z$-direction,
hence we focus our attention only on this component, unless stated
otherwise.  Based on the azimuthal velocity of the gas shown in the
right column of Fig.~\ref{fig:core_evolution}, we do not expect the
star to behave as a solid body, hence we divide it into shells where
we compute different diagnostic quantities.  For a rigid body, the
angular frequency and the total angular momentum are related by
$L_z=I_z \omega_z$, with $I_z$ being the moment of inertia with
respect to the $z$-axis.  Thus, for each shell we can estimate the
average angular frequency as
\begin{equation}
\langle\omega_z\rangle(r) = \frac{L_z}{I_z},
\end{equation}
with
\begin{equation}
I_z(r) = \sum_i m_i r_{\perp,i}^2,
\end{equation}
and
\begin{equation}
L_z(r) = \sum_i m_i r_{\perp,i} v_{\phi,i},
\end{equation}
where $r_{\perp,i}$ is the perpendicular distance to the star's
rotation axis, and $v_{\phi,i}$ is the azimuthal velocity. Notice that
the summation goes over all cells inside each particular shell.

We show the star's rotation during its interaction with the SMBH in
Fig.~\ref{fig:omega}.  The top panel of this figure shows the
evolution of the angular frequency profile, while the bottom panel
gives the cumulative angular momentum profile.  The angular frequency
shows strong evolution, starting roughly with rigid-body rotation
(yellow line). As the outer layers of the star are stripped, the core
spins up, departing greatly from a rigid-body.  Notice that for these
profiles we have replaced the radial coordinate by the enclosed mass,
because once the star is disrupted, the gas changes dramatically in
extent.  Recall that approximately half of the star survives for this
impact parameter, hence $m(<r)\sim 0.5\,{\rm M}_\odot$ corresponds
to the remnant's outer edge.

Roughly one day after the encounter, the remains of the star approach
an equilibrium where the outer layers rotate fast, while the inner
core is counter-rotating.
{It is important to clarify, however, that this result does not imply
that the inner core counter-rotates as a whole, but rather that its total
angular momentum is negative. As can be seen in the right column of
Fig.~\ref{fig:core_evolution}, the only material that has negative
azimuthal velocity is inside the two vortices, and thus this result
shows that these structures dominate the angular momentum
budget.}
By analysing the angular momentum of the
gas (Fig.~\ref{fig:omega}, lower panel) it is clear that during this
period there are no net torques acting at $m(<r)\sim 0.5\,{\rm M}_\odot$,
since the total value remains roughly constant. Consequently, the
evolution observed inside the core comes from the re-accommodation of
the different layers, which in turn re-distributes the angular
momentum.

\subsection{Vortex identification}

Due to the differential rotation inside the surviving core and with
respect to the surrounding material, there are clear vortices induced
in the fluid. In particular, about 2 hours after periapsis, two
prominent and persistent vortices develop inside the star. These
structures rotate together with the surviving core and clearly seem to
have an impact on the dynamics of the innermost region of the remnant.

In order to directly relate the presence of these two vortices to the
observed evolution, we need a method to identify their position in
every output.
{Because the instantaneous stream line pattern of a fluid can be
reconstructed through its velocity gradient, we can characterize
the motions inside the star using the eigenvalues of the velocity
gradient tensor $A$ \citep{Chong1990}.
This tensor is defined as $A_{ij}=\pdv*{v_i}{x_j}$,
and each gradient component is directly computed by \arepo~for each
cell at every timestep during the hydrodynamical run
\citep{Pakmor2016}.
Following \citet{Chong1990}, we expect the fluid to describe closed
or spiralling orbits (such as the ones expected in vortices)
if two of the eigenvalues of $A$ form a complex conjugate pair.
We define a vortex core as
the region where the velocity gradient tensor $A$ has complex
eigenvalues.}

As described by \citet{Chong1990}, the characteristic
equation for $A$ is given by
\begin{equation}
\lambda^3+\lambda^2P+Q\lambda+R=0,
\label{eq:lambda}
\end{equation}
where $P$, $Q$, and $R$ are the three invariants of $A$,
\begin{equation}
\begin{split}
P&= -\trace(A),\\
Q&= \frac{1}{2}\qty[\trace(A)^2-\tr(A^2)],\\
R&= -\det(A).
\end{split}
\end{equation}
The discriminant of equation~\eqref{eq:lambda} can be written as
\begin{equation}
\Delta \equiv 27R^2+(4P^3-18PQ)R+(4Q^3-P^2Q^2).
\end{equation}
The velocity gradient tensor will then
have one real eigenvalue and a pair of conjugated complex eigenvalues
whenever $\Delta>0$, and consequently we can use this condition to
determine which cells belong to stream lines resembling vortices.

Additionally, to further characterise these vortices, we use the fact
that the pair of complex eigenvalues can be written as
$\lambda_{cr}\pm i \lambda_{ci}$, where $\lambda_{ci}$ is usually
referred to as ``swirling strength'', as it is a measure of the local
swirling rate inside the vortex \citep{Zhou1999, Chakraborty2005}.
Hence, for each gas cell satisfying the $\Delta>0$ condition, we use
the imaginary part of its complex eigenvalues to quantify the strength
of the spiralling motions inside the star.  Finally, in order to
capture vortices in approximately closed orbits, we consider gas
cells that satisfy the conditions $\lambda_{ci}>\epsilon$ and
$-\kappa\leq \lambda_{cr}/\lambda_{ci}\leq \delta$, where $\epsilon$,
$\delta$, $\kappa$ take on non-negative values
\citep{Chakraborty2005}.  We find that
$\epsilon=5\times10^{-4}\,{\rm s}^{-1}$ and $\delta=\kappa=0.1$ are
appropriate to capture the vortices observed in our simulations.

In Fig.~\ref{fig:vortices} we show the motions in the stellar
interior, 4 hours after periapsis, where a vortex pair is clearly
present.  The left-hand panel of this figure displays the stream lines
of the fluid in the reference frame moving with the core.  The centre
panel shows $\lambda_{ci}$ of these cells. We can see clearly that the
strength is higher for the gas inside the vortices, with the peak
located on the rotation axis of both structures.

Finally, in order to estimate the position of the vortex pair in each
output, we compute the radial profile of the swirling strength inside
the star. An illustrative example is shown in the right-hand panel of
Fig.~\ref{fig:vortices}.  The profile displays the density of
$\lambda_{ci}$, namely, for each spherical shell we sum $\lambda_{ci}$
of the contained cells and normalize by its total volume.
We note that the presence of the vortex pair produces
a very prominent asymmetric peak.
However, since the fluid is highly turbulent within the surviving core,
this profile can be rather noisy in some of the outputs, mainly
due to the transient appearance of smaller vortices during
the evolution.
Consequently, we distinguish the main peak of each profile by fitting
a skew-normal distribution, which can be expressed as
\begin{equation}
f(x) = \frac{c}{\sqrt{2\pi}}e^{-\frac{(x-a)^2}{2b^2}}
\qty[1+\erf\qty(\frac{\alpha x}{\sqrt{2}})],
\end{equation}
where $a$, $b$, $c$, and $\alpha$ are the fitting parameters, and
$\alpha$ represents the ``skewness'' of the function.  Note that for
$\alpha=0$ this distribution is equivalent to a normal distribution.
We find this function appropriate to approximate the asymmetric
shape of the main peak, as seen in the example
displayed in the right-hand side panel of Fig.~\ref{fig:vortices}
with the dashed red line.
We assign the position of the vortex pair as the radius of the
maximum value of the fitted function.
The radial position of the vortex pair is shown with a dotted line
on each panel of Fig.~\ref{fig:vortices}, and it is clear that this
procedure yields a value that is consistent with the rotation
axis of each vortex.

\begin{figure}
\centering
\includegraphics[width=\columnwidth]{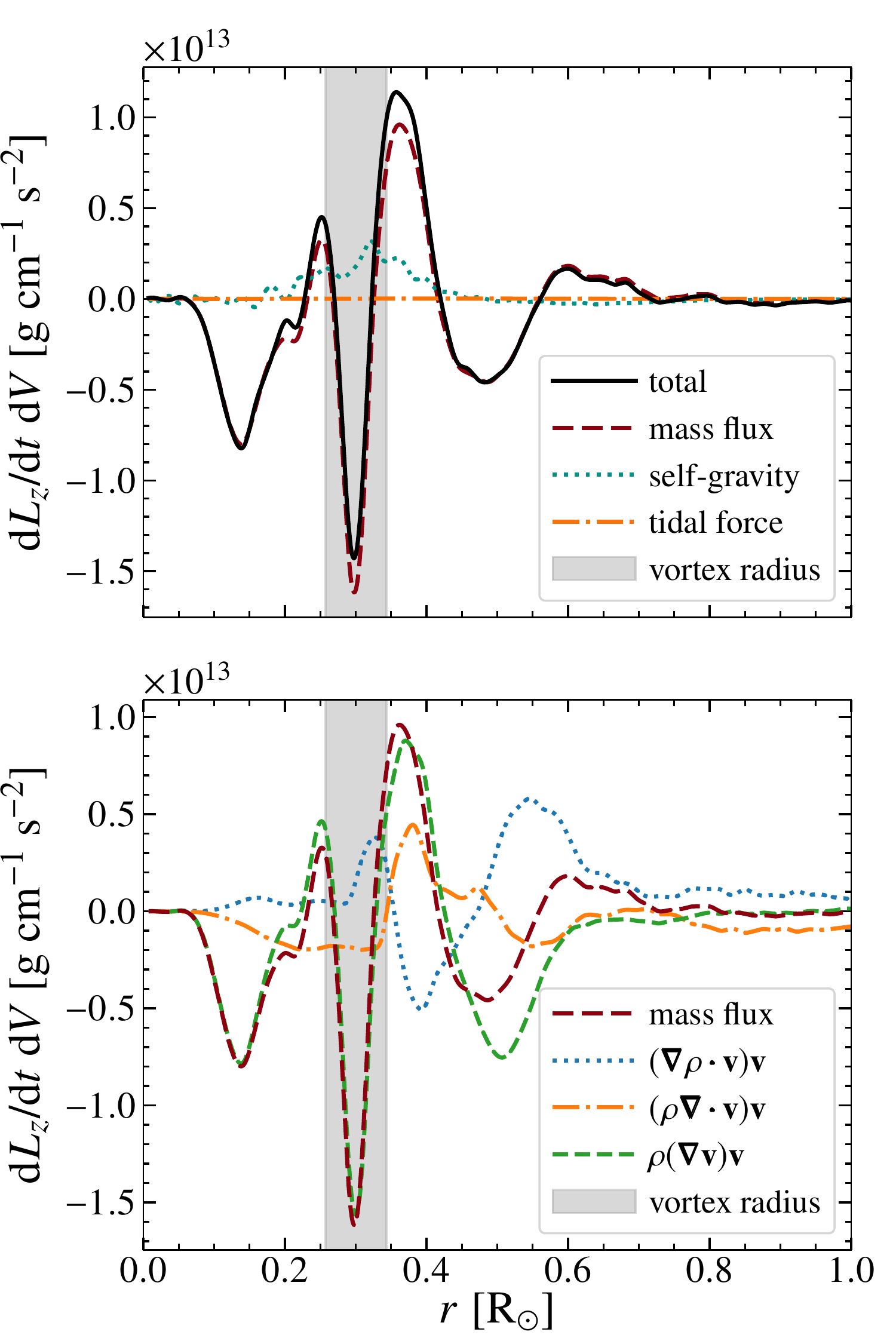}
\caption{Azimuthally averaged torque profile inside the star, averaged
  over the time interval $t=3-25$ h.  \textit{Upper panel}: Individual
  contributions of the different terms to the total angular momentum
  evolution (solid line) as shown in equation~\eqref{eq:torques}: mass
  flux (dashed line), gas self-gravity (dotted line) and BH gravity
  (dotted-dashed line).  \textit{Lower panel}: Decomposition of the
  mass flux term of the upper panel as shown in
  equation~\eqref{eq:mass_flux}.  }
\label{fig:torques}
\end{figure}

\subsection{Torque decomposition}

As observed in Fig.~\ref{fig:omega}, the rotation of the star evolves
greatly after periapsis.  In order to differentiate the forces
responsible for this evolution, and to establish the possible role of
the vortex pair, we compute the different torques inside the surviving
core.

The total angular momentum of a spherical shell can be written as
\begin{equation}
\vb{L}_{\rm shell}=\int_{\rm shell}\vb{r}\cross(\rho\vb{v})\dd V,
\end{equation}
thus its time derivative is
\begin{equation}
\pdv{\vb{L}_{\rm shell}}{t}=\int_{\rm shell}\vb{r}\cross\pdv{(\rho\vb{v})}{t}\dd V,
\label{eq:int1}
\end{equation}
which gives us the total torque over the shell.
From the Euler equation we have
\begin{equation}
\pdv{(\rho\vb{v})}{t}=-\div{}(\rho \vb{v}\vb{v}^T)-\grad P-\rho\grad\phi,
\label{eq:euler}
\end{equation}
{which inserted into equation~\eqref{eq:int1} gives us the different
contributions to the momentum evolution of the shell.}

{Intuitively, one would expect the pressure term not to contribute
to the total torque on each shell, since the pressure force points
always radial over the spherical surfaces we are considering.
To demonstrate that this is the case, we use the identity
$\curl(P\vb{r})=P \curl\vb{r}+\grad P\cross \vb{r}$ to yield
\begin{equation}
\int_{\rm shell} \vb{r}\cross\grad P\dd V = -\int_{\rm shell} \curl(P\vb{r})\dd V.
\end{equation}
Using the Green-Gauss theorem, we can transform the integral over
the volume of the shell to one over the enclosing surface
\begin{equation}
\int_{\rm shell} \curl(P\vb{r})\dd V = \int_{\rm surface} P(\vu{n}\cross\vb{r})\dd S,
\label{eq:int2}
\end{equation}
where $\vu{n}$ is the outward pointing unit normal vector to the shell's
surface.
Since by definition the normal vector to a spherical surface is radial
(i.e., $\vu{n}\equiv \vu{r}$),
the right-hand side of equation~\eqref{eq:int2} vanishes, which
is what we want to demonstrate.}

Finally, we discretise equation~\eqref{eq:int1} by summing
over the cells enclosed by each shell, which gives us an estimate of the torque
density profile
\begin{equation}
\frac{1}{\dd V}\dv{\vb{L}_{\rm shell}}{t}= -\frac{1}{\Delta V}\sum_i\vb{r}_i\cross \
\qty[\div{}(\rho_i \vb{v}_i\vb{v}_i^T)V_i
+ m_i \grad \phi_i],
\label{eq:torques}
\end{equation}
where
\begin{equation}
\Delta V = \sum_i V_i
\end{equation}
is the total volume of the cells contained in the shell.  The second
term on the right-hand side of equation~\eqref{eq:torques} is the
torque coming from the forces acting on the gas, which in this case
is only gravity. On the other hand, the first term can be interpreted
as the specific angular momentum flux between the different shells.

The gravitational potential can be split into contributions from
the gas self-gravity and the black hole potential.  However, it is
important to notice that the reference frame of the star is
non-inertial, which in practice means that the gravitational force
each gas parcel feels corresponds to a tidal stretching, which we can
be expressed as
\begin{equation}
\grad\phi_{\rm tidal} =  \frac{GM_{\rm BH}}{r_i^3}\vb{r}_i - \frac{GM_{\rm BH}}{r_*^3}\vb{r}_*.
\end{equation}

The top panel in Fig.~\ref{fig:torques} shows the different
contributions to the torque per unit {volume} inside the surviving core.
This radial profile is averaged over the time interval
$t=3-25\, {\rm h}$ after periapsis.  From this figure it is clear that
the torque is completely dominated by the redistribution of mass
inside the star, with a small contribution produced by the gas
self-gravity. As expected, the black hole gravity is negligible at
this stage, since the star is far from the tidal radius.  The total
torque has the largest (negative) value at $r\approx 0.3R_\odot$,
which coincides with the position of the prominent vortex pair
observed in our simulations.

The first term on the right-hand side of equation~\eqref{eq:euler} can
also be expanded to
\begin{equation}
\div{} (\rho \vb{v}\vb{v}^T) = (\grad\rho\vdot \vb{v})\vb{v}+(\rho\div\vb{v})\vb{v}+\rho(\grad\vb{v})\vb{v}.
\label{eq:mass_flux}
\end{equation}
Notice that $(\grad\vb{v})\vb{v}$ is a matrix vector multiplication,
as $\grad\vb{v}=\pdv*{v_i}{x_j}$ corresponds to the gradient velocity
tensor.  We show the contribution of the different terms from
equation~\eqref{eq:mass_flux} in the lower panel of
Fig.~\ref{fig:torques}.  We find that the dominant source of torque is
the term given by the velocity gradient, mainly because the other two
terms tend to cancel each other out.  Since $\grad \vb{v}$ is related
to the vorticity of the fluid, this suggests that the vortex pair is
responsible for the angular momentum transport inside the star, which
results in the counter-rotating core observed in Fig.~\ref{fig:omega}.
{To additionally support this mechanism, we can compare
the amount of angular momentum transported from the inner
core ($\tau_{\rm trans}$) with the value of the torque
measured in our simulation ($\tau_{\rm vort}$).
From Fig.~\ref{fig:omega} we have that initially the inner core
has a total angular momentum of $\sim$ 10$^{49}$~g~cm$^{2}$~s$^{-1}$,
which is transported in the span of $\approx$20 h
$\approx$ 7$\times$10$^4$ s.
This yields $\tau_{\rm trans}\sim$10$^{44}$~g~cm$^{2}$~s$^{-2}$.
On the other hand, from the torque density in Fig.~\ref{fig:torques}
we obtain $\tau_{\rm vort}$$\sim$~
10$^{13}$~g~cm$^{-1}$~s$^{-2}$~$\times$~$4\pi$~(0.3R$_\odot$)$^2$~
$\times$~0.1R$_\odot$ $\sim$ 10$^{44}$~g~cm$^{2}$~s$^{-2}$,
where the total vortex volume corresponds to the shaded area in
Fig.~\ref{fig:torques}.
Because these two torques are comparable, this estimation
strengthens our conclusion that the vortices produce the evolution
observed in the surviving core.}
We note that other examples of vortices responsible for angular
momentum transport have been found in the context of hydrodynamical
simulations of protoplanetary discs, where these spiralling motions
are long-lived structures that drive compressive motions
\citep[e.g.][]{Johnson2005}.

We find that the vortices are quite persistent, being still
present when we stop the simulations after several dynamical times of
the core.  This stability is expected given that the numerical
viscosity is very low, and we are not including physical viscosity
because non-perturbed stars can be approximated as inviscid given
their extremely high Reynolds numbers
\citep[see][ and references therein]{Miesch2009}.
However, for stellar cores that are the result of tidally disrupted
stars this might not necessarily be the case, especially if there are
strong magnetic fields. As shown by the numerical simulations of
\citet{Guillochon2017} and \citet{Bonnerot2017},
the vortex formation after the disruption
significantly amplifies the magnetic field inside the surviving core,
which can be a source of high viscosity. In fact, the vortices in
their simulations disappear after a few dynamical times, indicating
that the magnetic field is dissipating their energy.  Consequently, in
order to estimate a realistic dissipation timescale of the vortex pair
formed within the core, the addition of magnetic fields is crucial.

\section{Conclusions}
\label{sec:conclusions}

In this paper, we have introduced a new suite of simulations to study
the tidal disruption of stars by supermassive black holes, using for
the first time the hydrodynamical moving-mesh code \arepo. This code
has been previously used to investigate a large number of
astrophysical problems in the areas of cosmic structure formation and
galaxy evolution, and it has also been used in a few selected
applications in stellar astrophysics. We have here shown that it also
provides a powerful tool to study TDEs with unprecedented
accuracy. This is because \arepo~can accurately follow high-speed
orbits while still well resolving mixing and shocks in the rest-frame
of the moving fluid. This combination is usually not readily available
with more standard numerical techniques, because SPH is comparatively
noisy whereas Eulerian grid-based codes are diffusive for the high
bulk velocities occurring in the TDE problem.

Since the appearance of flares produced by TDEs in the core of
galaxies depends critically on whether the star is fully or partially
destroyed, we simulated a total of 15 encounters with different impact
parameters in order to determine the threshold between these regimes.
As demonstrated by \GRR, the critical distance for total disruption
depends strongly on the stellar structure. Hence it is of paramount
importance to use physically motivated structures for stars in TDE
simulations to produce realistic predictions.
In this paper, we model the stellar structure using a $1\,{\rm M}_\odot$
zero-age main sequence profile obtained with the stellar evolution
code MESA, in contrast to the single polytrope that is often chosen
in this type of hydrodynamical simulations.

Our main results can be summarized as follows:
\begin{itemize}

\item We find that the star studied here is completely destroyed for
  penetration parameters $\beta\gtrsim 2$.

\item The mass loss of the star as a function of $\beta$ is similar to
  the one obtained with a 4/3 polytrope, including the value of the
  critical distance for total disruption. This is consistent with the
  fact that the latter is a decent approximation for our ZAMS profile.

\item As in previous works, we find that the shape of the energy
  distribution of the material stripped from the star depends on the
  fate of the star. Encounters resulting in a surviving core deplete
  the gas at lower energies.  The spread of the energy distribution is
  significantly larger than expected under the `frozen in'
  approximation, where the energy of each gas parcel is fixed at its value at the
  tidal radius.  This indicates that the internal forces of the star
  are able to redistribute some the energy, most likely through shocks
  close to periapsis.

\item Using the energy distribution of the gas still bound to the
  SMBH, we computed the fallback rates as a function of time, which
  provides an estimate of the SMBH mass accretion rate.  The
  gravitational influence of the surviving core causes deviations of
  the slope of this decay towards steeper values at late times with
  respect to the theoretical expectation ($\dot M\propto t^{-5/3})$.
  We found that only deep encounters $(\beta\gtrsim 2.5)$ result in
  fallback rates consistent with the expected decay.

\item The hydrodynamical evolution of a surviving core after a grazing
  encounter ($\beta=1.6$) is complex.  The vorticity of the fluid
  inside the core shows plenty of substructure, mainly in the form of
  small filaments, revealing a turbulent evolution of the star after
  the encounter. Furthermore, the gas velocity shows the presence of
  two prominent vortices on opposite sides of the core. As this pair
  rotates with the core, the fluid is constantly being stirred up,
  promoting turbulent behaviour.

\item  {The surviving core ends up with positive angular
  momentum in the outer layers, while negative in the innermost
  region. We found that}
  this configuration is achieved by the internal
  forces of the stellar core, rather than external torques. We also
  found that the strongest torques directly correlate with the
  location of the vortex pair, and that the vortex pair is largely
  responsible for the angular momentum transport inside the surviving
  stellar core.

\end{itemize}

It is apparent from the results described above that the evolution of
the stellar interior structure during the close encounter with a SMBH
is very rich. This could have some very interesting implications, and
opens up different avenues for future simulations using the
moving-mesh technique.  For instance, it is well known that
differential rotation in stellar interiors can dramatically change the
evolution of stars. For instance, shear instabilities triggered by
differential rotation can generate turbulence, and hence induce extra
mixing. This rotational mixing plays an important role in the
evolution of massive stars \citep[e.g.][]{Zahn2004, Zahn2008}.  Thanks
to the grid-based scheme of \arepo, mixing is naturally
resolved. Consequently, simulations as the ones presented in this
paper could be used to study the redistribution of material inside the
star. In particular, we could track the composition of the gas inside
the surviving core, which can be achieved by using tracer particles
that can follow the fluid in a Lagrangian way \citep{Genel2013}.  Once
the stellar remnant reaches an equilibrium, the resulting distribution
could be mapped to a one dimensional profile and further evolve using
a stellar evolution code such as MESA.  This procedure has the
potential to reveal some unique observational features from stellar
remnants from a close encounter with a SMBH.  If such a population of
unusual stars was to be discovered in our Galactic Centre or local
galaxies with future surveys, this could help constrain the rate of
TDE in such galaxies, which in turn constrains the dynamical nature of
the nuclear stellar core \citep{Frank1976,Syeer1999,Wang2004}.

Additionally, mixing could be studied within the material stripped
from the star.  As suggested by \citet{GallegosGarcia2018},
compositional changes resulting from the fallback gas could be
reflected in the TDE light curve, as well as the spectra. These
features could be used to constrain the properties of the disrupted
star. However, as these authors acknowledge, their simple framework
should only be taken as a guide, and hydrodynamical simulations along
the lines carried out here are needed to overcome the approximations
employed in these previous estimates.

A further advantage of our approach is that
it can be extended to any stage of stellar evolution, adopting physically
motivated initial profiles obtained from stellar evolution codes.
This opens up the possibility of
modelling tidal disruptions for a whole suite of different stars, with
varying mass and age, but always using a realistic internal structure. This
includes giant stars, for example, where we replace the core with a point
mass to represent the extreme density contrast between the core and
the envelope \citep{Ohlmann2017}.
While main-sequence stars are swallowed whole by high mass
black holes ($\gtrsim 10^8$M$_\odot$) producing almost no emission
\citep{MacLeod2016},
giant stars could be used to probe the higher end of the black hole
mass function.
However, as suggested by \citet{Bonnerot2016b}, following the disruption
of giant stars, the interaction between the debris stream and the gaseous
environment could reduce the amount of gas available for accretion.
Using analytical arguments, the authors show that Kelvin-Helmholtz (K-H)
instabilities can affect the debris, dissolving a substantial fraction
of the stream still bound to the black hole.
This type of problem is well suited for \arepo, since its treatment
of contact discontinuities retains the necessary accuracy to resolve
instabilities such as K-H.

Finally, it is clear from our simulations that the early evolution of the
surviving core is largely dominated by the presence of the vortex
pair.  As this pair rotates with the stellar remnant, it will likely
promote turbulence and mixing, as well as magnetic field
amplification.  It will be interesting to examine these aspects in
future simulations, in addition to study the stellar evolution of
partially disrupted stars, which potentially may be reflected in
peculiar observational signatures.

\section*{Acknowledgements}

{We are grateful to the anonymous referee for very insightful
corrections and suggestions that helped improve substantially
the clarity of our paper.}
The left-hand side plot in Fig.~\ref{fig:vortices} was produced using
the python packages from the yt project
(\url{https://yt-project.org/}).  The simulations were performed on
the computer cluster at HITS.  FGG acknowledges support from the
Deutscher Akademischer Austauschdienst DAAD (German Academic Exchange
service) in the context of the PUC-HD Graduate Exchange Fellowship.
This work was partially supported by the European Research Council
under ERC-StG grant EXAGAL-308037, and the Klaus Tschira Foundation.




\bibliographystyle{style/mnras}
\bibliography{bib/refs} 



%
%


\bsp	
\label{lastpage}
\end{document}